\documentclass[aps,prb,twocolumn,superscriptaddress,longbibliography]{revtex4-2}

\usepackage{mathrsfs}
\usepackage{tabularx}
\usepackage{slashed}
\usepackage{amsmath}
\usepackage{amsfonts}
\usepackage{bm}
\usepackage{graphicx,color}
\usepackage{times}
\usepackage{braket}
\usepackage{orcidlink}
\usepackage[caption=false]{subfig}
\AtBeginDocument{%
   ~\newwrite\bibnotes
   ~\def\bibnotesext{Notes.bib}
   ~\immediate\openout\bibnotes=\jobname\bibnotesext
   ~\immediate\write\bibnotes{@CONTROL{REVTEX41Control}}
   ~\immediate\write\bibnotes{@CONTROL{%
    apsrev42Control,author="08",editor="1",pages="1",title="0",year="1"}}
    ~\if@filesw
    ~\immediate\write\@auxout{\string\citation{apsrev42Control}}%
   ~\fi
}%

\begin{document}

\title{Exciton fractional Chern insulators in moiré heterostructures}

\author{Raul Perea-Causin  \orcidlink{0000-0002-2229-0147}}\thanks{raul.perea.causin@fysik.su.se}
\author{Hui Liu \orcidlink{0009-0009-4988-9561}}\thanks{hui.liu@fysik.su.se}
\author{Emil J. Bergholtz \orcidlink{0000-0002-9739-2930}}\thanks{emil.bergholtz@fysik.su.se}
\affiliation{Department of Physics, Stockholm University, AlbaNova University Center, 106 91 Stockholm, Sweden}

% \date{\today}

\begin{abstract}
Moiré materials have emerged as a powerful platform for exploring exotic quantum phases. While recent experiments have unveiled fractional Chern insulators exhibiting the fractional quantum anomalous Hall effect based on electrons or holes, the exploration of analogous many-body states with bosonic constituents remains largely uncharted. In this work, we predict the emergence of bosonic fractional Chern insulators arising from long-lived excitons in a moiré superlattice formed by twisted bilayer WSe$_2$ stacked on monolayer MoSe$_2$. Performing exact diagonalization on the exciton flat Chern band present in this structure, we provide compelling evidence for the existence of Abelian and non-Abelian phases at band filling $\frac{1}{2}$ and $1$, respectively, through multiple robust signatures including ground-state degeneracy, spectral flow, many-body Chern number, and particle-cut entanglement spectrum. The obtained energy gap of $\sim 10$ meV for the Abelian states suggests a remarkably high stability of this phase, which persists for a relatively wide range of twist angles and vertical electric fields. Our findings establish the presence of robust bosonic fractional Chern insulators in highly tunable and experimentally accessible moiré heterostructures and unveil a promising pathway for realizing non-Abelian anyons.
\end{abstract}

\maketitle

Fractional Chern insulators (FCIs)---lattice analogs of the fractional quantum Hall effect that remain robust in the absence of a magnetic field---hold large potential for the study of fundamental quantum phenomena and for the development of novel quantum technologies~\cite{fqahe_review,PARAMESWARAN2013816,Zhao_Liu_Review}.
Pioneering experiments~\cite{spanton2018observation,xie2021fractional} and predictions~\cite{ahmed_fci,repellinChernBandsTwisted2020,Ledwith_analytical_approach_FCI} of FCIs in twisted van der Waals heterostructures, followed by recent realizations at absent magnetic field~\cite{FCI_MoTe2_3,FCI_MoTe2_2,PhysRevX.13.031037,lu2024fractional}, have established moiré materials as an accessible and versatile platform for exploring strongly-correlated topological phases.
This breakthrough has stimulated abundant efforts in the pursuit of exotic phases beyond the conventional paradigm of Laughlin and hierarchy fractional quantum Hall states.
In particular, predictions of Moore--Read (MR)~\cite{mr_Aidan, mr_Ahn, mr_Donna, mr_Hui, mr_Xu, mr_Wang}
and Read--Rezayi~\cite{our_parafermion} phases hosting non-Abelian anyon excitations are especially promising for fault-tolerant topological quantum computing~\cite{RevModPhys.80.1083}. So far, however, research on moiré FCIs has focused only on the approach of doping the system with electrons or holes---resulting in correlated topological phases with fermionic constituents and leaving their bosonic counterpart largely unexplored.

Excitons, i.e. Coulomb-bound pairs of  conduction-band electrons and valence-band holes~\cite{excitons_colloquium}, are obvious candidates for realizing correlated bosonic phases in moiré materials~\cite{semiconductor_moire_review,exciton_TMDbilayers_review}. Concretely, interlayer excitons with charge-carriers located in different layers are particularly promising due to their long lifetime, which can reach hundreds of nanoseconds~\cite{jauregui_interlayer,elaine_lifetime} and even microseconds~\cite{excitonlifetime_microsecond}. These species typically appear in semiconducting van der Waals heterostructures, where an optical excitation generating intralayer excitons (with electrons and holes in the same layer) is shortly followed by tunneling of either electrons or holes into another layer~\cite{giuseppe_nature}.
The resulting interlayer excitons possess a permanent dipole moment, which makes them highly tunable by an out-of-plane electric field~\cite{mak_interlayer_excitons,tagarelli_natphot}.
Moreover, repulsive dipolar interactions together with the presence of flat bands in moiré semiconductors lead to correlated exciton physics ~\cite{correlated_exciton_insulator_natphys,dipolar_exciton_insulator_natphys,exciton_insulator_natphys,correlated_exciton_insulator_science}.
In addition, the large degree of control via twist angle, dielectric environment, and electric field can be exploited to achieve topological bands~\cite{Fengcheng_intralayer_exciton}.
In particular, long-lived interlayer excitons in a moiré heterostructure consisting of twisted bilayer WSe$_2$ on monolayer MoSe$_2$ (tWSe$_2$--MoSe$_2$), which exhibit topological flat bands, have been proposed as a promising route for the exploration of exciton physics with intertwined correlations and topology~\cite{dasSarma_topologicalExcitons}. 

Crucially, however, the emergence of FCI topological order is not at all guaranteed in a topological flat band, and even less so in the case of non-Abelian states which usually require artificially adjusted interactions.
For instance, in order to emulate the lowest Landau level and thus host Laughlin zero modes in presence of short-range pseudopotential interactions, a band must have an ideal quantum geometry (or be vortexable) in addition to being flat and topologically non-trivial~\cite{PhysRevB.90.165139,jie_wang_qgt2,Vortexability_band}. Even such ideal bands host non-FCI states in some cases~\cite{Hui_broken_symmetry,raulQAHC,ji2024quantummetricinducedhole}, and, all the more surprising, FCI order can emerge in trivial bands~\cite{simonFCIzeroBerry}. Thus, a topological flat band by itself is definitely not a proof of FCI order, and the presence or absence of topological order must be addressed with many-body calculations.

\begin{figure}[t]
\centering
\includegraphics[width=\linewidth]{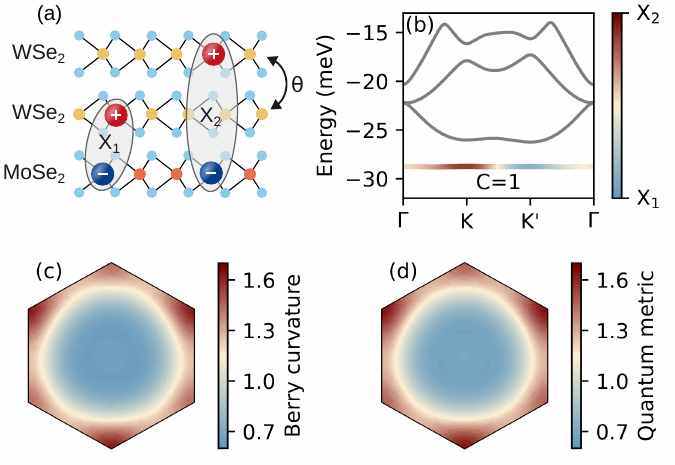}
\caption{Nearly ideal exciton Chern band. (a) Schematic illustration of the considered tWSe$_2$--MoSe$_2$ structure. The interlayer excitons X$_1$ and X$_2$ are formed by an electron in MoSe$_2$ and a hole in either of the two WSe$_2$ layers.
(b) Exciton band structure for $\Delta=-3.8$ meV and $\theta=1.95^{\circ}$, where the lowest band has a Chern number $C=1$. The color represents the contribution from each exciton species to the band.
(c) Berry curvature $\Omega_\mathbf{k} A_\text{BZ}/2\pi$ and (d) Fubini-Study metric $\text{tr}[g_\mathbf{k}] A_\text{BZ}/2\pi$ of the flat band across the moiré Brillouin zone. $A_\text{BZ}$ is the Brillouin zone area.
}
\label{fig:band}
\end{figure}

In this work, we unveil the many-body topological exciton phases emerging in a moiré heterostructure.
First we show that, besides being a nearly-flat Chern band, the lowest exciton band in tWSe$_2$--MoSe$_2$ exhibits an almost ideal quantum geometry---posing this system as a strong candidate for realizing exciton FCI phases.
Employing exact diagonalization, we show that the ground state at half filling with contact interactions is analogous to the bosonic Laughlin state in the lowest Landau level. Concretely, the state is characterized by a twofold degeneracy as well as an approximately zero energy, and its nature is further confirmed by the many-body Chern number, spectral flow, and entanglement spectrum.
Importantly, the Laughlin states remain robust and exhibit a large gap ($\sim 10$ meV) when replacing the idealized contact potential by realistic long-range interactions.
Furthermore, the gap remains sizable across a wide range of experimentally accessible twist angles and electric fields, where the latter controls the energy offset between the two interlayer excitons (X$_1$ and X$_2$ in Fig.~\ref{fig:band}(a)).
Finally, the many-body calculations at filling $1$ provide compelling evidence for a stable bosonic version of the non-Abelian MR state.
Overall, our findings predict the existence of exciton FCIs in an experimentally accessible moiré heterostructure and pave the way for the realization of Abelian and non-Abelian topological phases with bosonic constituents in moiré materials.

\emph{Nearly-ideal exciton Chern band---}%
We consider the van der Waals heterostructure tWSe$_2$--MoSe$_2$, where an optical excitation and subsequent tunneling result in the formation of long-lived interlayer excitons composed of a hole in either of the two WSe$_2$ layers and an electron in MoSe$_2$, cf. Fig.\,\ref{fig:band}(a).
The formation and equilibration of these excitons occurs on an ultrafast sub-picosecond equilibration time~\cite{giuseppe_nature}---many orders of magnitude smaller than the (up to microsecond~\cite{excitonlifetime_microsecond}) lifetime.
Interlayer excitons can also be obtained directly by electrical doping of the electron and hole layers~\cite{exciton_insulator_nat} typically separated by an hBN spacer.
We assume spin-valley polarization of electrons and holes, which can be achieved by a circularly-polarized optical excitation if the valley decoherence time is sufficiently long and, in some cases, occurs spontaneously due to interactions.

In order to describe moiré excitons in this structure, we consider the model introduced in Ref.~\cite{dasSarma_topologicalExcitons} and outlined in the Supplemental Material (SM)~\cite{SupMat,Fukui_Chern,Repellin_tilted_sample}.
In particular, we diagonalize the moiré exciton Hamiltonian~\cite{dasSarma_topologicalExcitons,brem_moire} $H_\text{x,m}$, which describes how the two relevant interlayer exciton states are affected and coupled by the moiré potential and tunneling that holes in the WSe$_2$ layers experience.
The lowest band has a Chern number $C=1$ for a specific range of the twist angle and the energy offset between the two exciton species, $\Delta=E^\text{x}_2-E^\text{x}_1$. The latter can be experimentally controlled by an out-of-plane electric field.

Here, we consider the twist angle $\theta=1.95^{\circ}$ and the offset $\Delta=3.8$ meV where, besides possessing a finite Chern number $C=1$, the lowest band is flat, cf. Fig.~\ref{fig:band}(b).
Interestingly, the quantum geometry of the band is nearly ideal~\cite{PhysRevB.90.165139,jie_wang_qgt2}, i.e. $\mathrm{tr}[g_\mathbf{k}] \approx |\Omega_\mathbf{k}|$ where $§g_\mathbf{k}$ is the quantum (Fubini--Study) metric~\cite{wojciech2025quantummetric} and $\Omega_\mathbf{k}$ is the Berry curvature, cf. Fig.~\ref{fig:band}~(c),~(d). This property strongly suggests the emergence of zero-energy ground states at even-denominator filling of the band with pseudopotential interactions---in analogy to Laughlin states of bosons in Landau levels, albeit now in the absence of a magnetic field. A nearly ideal quantum geometry was also found to be the precursor of electron FCIs in twisted bilayer graphene~\cite{Ledwith_analytical_approach_FCI,ideal_band_tbg} despite the fact that fluctuations of the metric induce new competing states \cite{PhysRevResearch.5.L012015}.

\begin{figure*}
\centering
\includegraphics[width=0.9\linewidth]{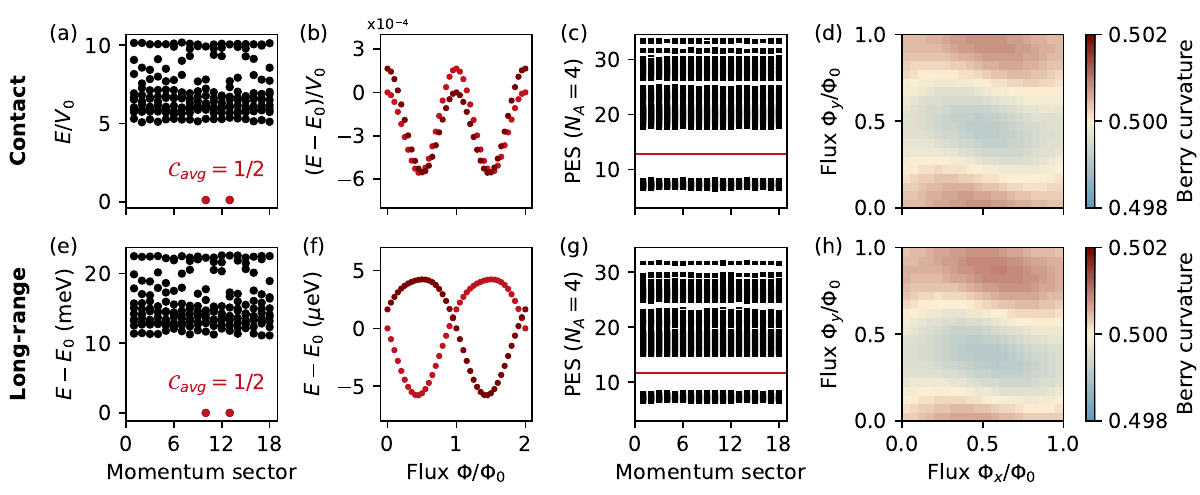}
\caption{Laughlin states at half filling. (a) Many-body energy spectrum containing the 10 lowest energies for each momentum sector, (b) spectral flow, (c) particle-cut entanglement spectrum, and (d) many-body Berry curvature for the two ground states considering contact interactions without kinetic energy effects at $\nu=\frac{1}{2}$. The respective data considering the realistic long-range interaction and the kinetic energy is shown in (e)-(h). The ground states in (a) and (e) are marked in red and have an average many-body Chern number $C_\text{avg}=1/2$. In the PES, the number of states below the first entanglement gap (denoted by the red solid line) is $1287$, matching the number of quasi-hole excitations in the $\nu=\frac{1}{2}$ Laughlin states. The considered system has $N_\text{s}=18$ moiré sites.}
\label{fig:Laughlin}
\end{figure*}

In order to elucidate whether the nearly ideal exciton Chern band indeed hosts FCI order, we will exactly diagonalize the many-body Hamiltonian $H_\text{x} = H_\text{x,m} + H_\text{x-x}$ projected into the lowest band. 
The interaction Hamiltonian, $H_\text{x-x}$, is described in the End Matter.
We consider finite-sized systems consisting of $N_\text{x}$ excitons in $N_\text{s}$ moiré sites with the filling factor $\nu=N_\text{x}/N_\text{s}$.
Importantly, the exciton operators $X^{(\dagger)}_\mathbf{Q}$ obey bosonic commutation relations. Corrections to the bosonic character arising from the exciton's fermionic substructure are typically expected in the regime where the distance between neighboring excitons is comparable to the size of an exciton~\cite{HAUG19843}, i.e. $n_\text{x}a_\text{B}^2 \sim 1$  where $n_\text{x}$ is the exciton density and $a_\text{B}$ the exciton Bohr radius. In typical moiré systems where there is one exciton per moiré site, $n_\text{x}a_\text{B}^2 \sim 0.01$, the bosonic description is appropriate.

\emph{Abelian states at half filling---}%
Based on the criteria for how closely a moiré band can mimic a Landau Level---(i) nontrivial topology, (ii) flat dispersion, and (iii) ideal quantum geometry~\cite{PhysRevB.90.165139,Zhao_Liu_Review,jie_wang_qgt2}---the exciton flat band shown in Fig.~\ref{fig:band}(b) appears to be an excellent candidate for realizing FCI phases analogous to the fractional quantum Hall effect of bosons. 
In this context, Laughlin states at half filling constitute a prototypical phase~\cite{Laughlin,haldane_fqhe}. Theoretically, they emerge as exact zero-energy states in a system of bosons with contact interactions in the lowest Landau level and their elementary excitations behave as Abelian anyons.
Bosonic Laughlin states have also been proposed in a different setting, where the bosons correspond to particle-hole excitations out of a fully-filled valley-polarized electron band~\cite{oxford_fci_exciton,sodemann_fci_exciton}---however, these states require fine-tuned interactions in order to be stable, the systems do not offer a way of controlling the filling factor, and predictive numerical calculations are lacking. Additionally, composite-fermion mean-field theory based on a generic BCS-like Hamiltonian suggested the potential emergence of Laughlin states from the topological p-wave pairing of electrons and holes in an excitonic insulator~\cite{Kane_fci_exciton}.
Here, we consider long-lived interlayer excitons in the nearly ideal, yet realistic, exciton Chern band displayed in Fig.~\ref{fig:band}(b) and provide numerical evidence showing that it supports remarkably stable Laughlin states even for realistic long-range interactions.

In order to strengthen the ideal conditions of the system and establish the emergence of Laughlin states in the exciton Chern band, we initially consider contact interactions and disregard the impact of the kinetic energy. With these assumptions, exact diagonalization yields two many-body ground states with approximately zero energy, cf. Fig.~\ref{fig:Laughlin}(a). Importantly, the ground states are separated from excited states by a large energy gap, which reflects the topological protection.
The twofold degeneracy and the total momentum of each state are characteristic of Laughlin states and can be understood with the aid of Landau level physics in the thin-torus limit~\cite{bergholtz2005half}. 
In particular, the thin-torus exclusion principle for $\nu=\frac{1}{2}$ Laughlin states dictates that a zero-energy state contains, at most, one particle in two consecutive sites~\cite{ardonne2008degeneracy}. The two states that fulfill this principle and therefore constitute the two degenerate ground states are those with a Fock-space configuration $101010\cdots$ and its translational-invariant partner $010101\cdots$, whose total momenta match those of our numerical ground states.
Furthermore, upon threading a magnetic flux (corresponding to twisted boundary conditions), the ground states evolve into each other and return to their original states after the insertion of one and two flux quanta (Fig.~\ref{fig:Laughlin}(b)), respectively, indicating that the two states are adiabatically connected and that the Hall conductivity is quantized to $\frac{1}{2}$.
The latter aspect is further confirmed by a direct calculation of the many-body Chern number, which yields $C_\text{avg}=\frac{1}{2}$ for each ground state and arises from a homogeneous many-body Berry curvature, cf. Fig.~\ref{fig:Laughlin}(d).
Next, we calculate and analyze the particle-cut entanglement spectrum (PES), which reveals the nature of the elementary quasi-hole excitations in the ground states (see SM). The calculated PES exhibits a large entanglement gap, below which the number of states matches exactly the analytical counting of quasi-hole excitations in the $\nu=\frac{1}{2}$ Laughlin states (Fig.~\ref{fig:Laughlin}(c)). Thus, the many-body ground state degeneracy, spectral flow, many-body Chern number, and PES clearly confirm the emergence of a stable FCI phase of excitons in the half-filled moiré Chern band.

We now consider realistic long-range exciton--exciton interactions (see End Matter) and take into account the impact of the small but finite kinetic energy.
The long-range interaction potential contains electron--electron, hole--hole, and electron--hole interaction mechanisms and is governed by the direct component, which at long distances behaves as the interaction between two dipoles with lengths $dl$ and $dl'$, i.e. $V_{ll'}(\mathbf{r}) \sim (dl)(dl')/|\mathbf{r}|^3$. Formally, the interaction potential also contains exchange terms, but typically these play a minor role in the context of interlayer excitons~\cite{erkensten_interlayerExcitons,steinhoff_xxinteractions,kyriienko_excitons,samuel_bosonic}.
Considering these realistic exciton--exciton interactions, the results of exact diagonalization and the subsequent analysis are remarkably similar to those obtained for contact interactions, cf. Fig.~\ref{fig:Laughlin}(e)-(h).
The fact that the energy gap between ground and excited many-body states ($\sim 10$ meV) persists for the considered systems ranging from $N_\text{s}=10$ up to $N_\text{s}=20$ sites clearly reflects that such phases are robust (see SM).
Thus, our numerical calculations demonstrate that exciton FCIs are stable in the tWSe$_2$--MoSe$_2$ moiré structure.

\begin{figure}
\centering
\includegraphics[width=\linewidth]{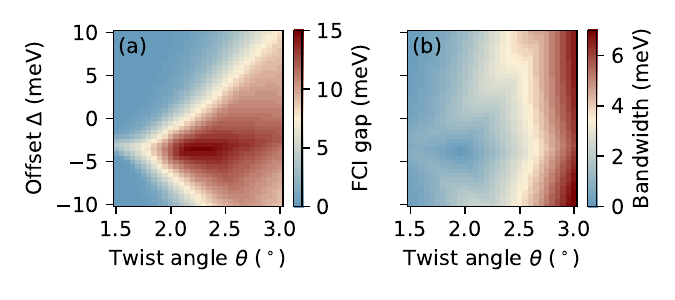}
\caption{Stability of Laughlin states in the $(\theta,\Delta)$ parameter space, where $\theta$ is the twist angle and $\Delta$ is the energy offset between the two interlayer exciton resonances.
(a) Energy gap in the many-body spectrum between the Laughlin ground states and excited states at filling $\nu=\frac{1}{2}$ in a system of $N_\text{s}=12$ sites with long-range interaction.
(b) Bandwidth of the lowest single-particle exciton band.
}
\label{fig:parameter_scan}
\end{figure}

Furthermore, we show that the many-body energy gap remains large ($\gtrsim 10$ meV) for a broad range of twist angles $\theta$ and energy offsets $\Delta$, cf. Fig.\ref{fig:parameter_scan}(a). In particular, the Laughlin-like FCI phase seems to be stable as long as the single-particle exciton band is topological (see SM), and a sizable gap persists even at large twist angles ($\theta \sim 3^\circ$) where the band is not flat anymore, cf. Fig.~\ref{fig:parameter_scan}(b).
Our calculations indicate that the $\nu=\frac{1}{2}$ exciton FCI phase is most robust at twist angles $\theta \in [2^\circ, 2.5^\circ]$ and offsets $\Delta \in [-5\ \text{meV}, 0\ \text{meV}]$, where the many-body and PES gaps are largest and the energy spread between the two ground states is smallest (see SM).

Interestingly, the calculated gap of $\sim 10$ meV is two times larger than in the case of hole FCIs in the similar system of twisted bilayer MoTe$_2$~\cite{PhysRevB.108.085117}. In principle, this suggests a higher stability of exciton FCIs compared to their electron/hole counterparts, whose gap has been experimentally estimated to be 20 K~\cite{xiaodong_high_temp_fci}.
The potentially larger gap of exciton FCIs might be a result of the stronger repulsive interaction dominated by $V^\text{e-e}+V^\text{h-h}$ in the short range ($V^\text{e-h}$ is weaker since the charges are located in separate layers) instead of just $V^\text{e-e}$ or $V^\text{h-h}$.
A larger gap for bosons can also be expected from a pseudopotential perspective, where the interaction for bosons and fermions is dominated by the zeroth $v_0$ and first $v_1$ pseudopotentials, respectively, where generally $v_n> v_{n+1}$ in lowest Landau level-like bands~\cite{haldane_fqhe,Ledwith_analytical_approach_FCI}.
We note, though, that a reliable quantitative estimation of the gap is difficult due to limitations such as the finite system size, the use of a pure Coulomb potential~\cite{excitons_colloquium}, and uncertainties in parameters such as the dielectric constant.

\emph{Non-Abelian states at filling one---}%
After confirming the presence of Abelian exciton FCIs, we now seek exotic phases whose elementary excitations obey non-Abelian anyon statistics.
In this context, the most straightforward and promising candidate is the MR state~\cite{MOORE1991362}.
This state can be theoretically understood as the exact zero-energy ground state at filling $\nu=1$ for bosons in the lowest Landau level with artificial three-body contact interactions~\cite{greiter1992} and it has been predicted to appear in rotating Bose-Einstein condensates~\cite{cooper_rotatingBEC,regnault_rotatingBEC}.
In principle, MR states can be stabilized by long-range two-body interactions~\cite{NAlong2013}, which are at least more favourable than (two-body) short-range pseudopotentials for achieving the required pairing. In addition, the bosonic quantum statistics might further facilitate the pairing and, thus, the stabilization of the typically elusive MR states.
In the following, we show numerical evidence indicating that this phase is stable in the realistic conditions considered here, i.e. in the moiré exciton band and assuming long-range interactions.

\begin{figure}
\centering
\includegraphics[width=\linewidth]{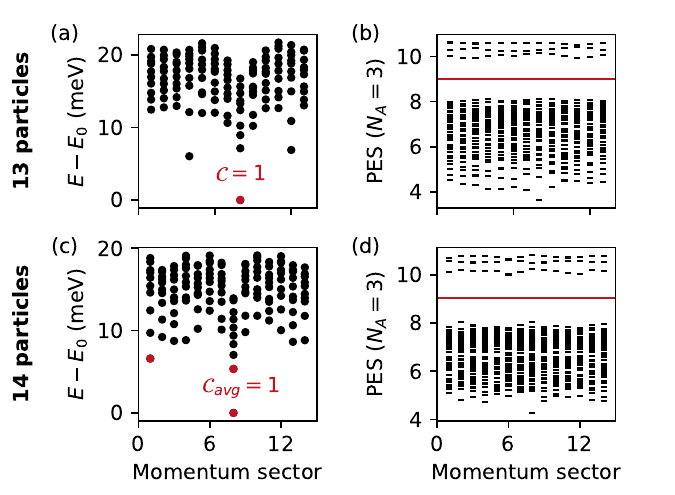}
\caption{Moore--Read states at filling one. Many-body spectrum (10 lowest energies) and PES at $\nu=1$ with long-range interactions for (a)-(b) 13 and (c)-(d) 14 particles. The single and threefold quasi-degenerate ground states (red dots in panels (a) and (c)) for an odd and even particle number of sites are located at the momenta expected for MR states. The red line in panels (b) and (d) indicates the entanglement gap respecting the quasi-hole counting rules of the MR state (416 and 518 quasihole excitations for 13 and 14 particles with $N_A=3$).}
\label{fig:MooreRead}
\end{figure}

First, we note that the most prominent feature of the MR phase is the dependence of the ground state degeneracy on the parity of the particle number. Here, the thin-torus exclusion principle~\cite{Bergholtz2006} states that, at most, two particles can occupy two consecutive orbitals~\cite{Seidel2006,ardonne2008degeneracy}, resulting in three possible Fock-space configurations, $111111\cdots$, $202020\cdots$, and $020202\cdots$. As a result of periodic boundary conditions, only the first configuration is allowed for an odd number of particles---resulting in a single ground state---, while for an even number of particles all three configurations are allowed and the ground state is threefold degenerate.

In Fig.~\ref{fig:MooreRead}(a),(c), we show the calculated many-body spectra for systems with 13 and 14 excitons. The single and three lowest states in the case of odd and even number of excitons, respectively, are located at the expected momenta for MR states.
We note that the spread of the ground state energies and the small gap with respect to excited states is a common feature of finite-size systems~\cite{mr_Aidan, mr_Ahn, mr_Donna, mr_Hui, our_parafermion}. We expect the states to become exactly degenerate in the thermodynamic limit, and we note that the gap persists upon threading magnetic flux and for different system sizes (see SM).
Moreover, the calculation of many-body Chern number yields $\mathcal{C}_{\text{avg}}=1$ for each state and reveals a uniform Berry curvature distribution, indicating that the MR phase is robust.

In order to unambiguously determine the nature of the phase beyond the parity dependence of the ground-state degeneracy, we calculate the PES. For both $N_\text{x}=13$ and $N_\text{x}=14$ systems, the PES is characterized by a large gap, below which the number of states exactly matches the counting of quasi-hole excitations allowed by the MR exclusion rule. Furthermore, the PES gap persists across various system sizes and different $N_A$ (see SM). All together, these results constitute convincing evidence for the presence of a robust non-Abelian MR phase with exciton constituents at filling $\nu=1$ in the tWSe$_2$--MoSe$_2$ moiré structure.

\emph{Conclusion and outlook---}%
We have explored the emergence of strongly-correlated topological phases arising from a bosonic system of long-lived moiré excitons in a flat Chern band. 
In particular, by combining exact diagonalization with many-body diagnosis tools, we have demonstrated the existence of robust Abelian and non-Abelian exciton fractional Chern insulators in a realistic model of the tWSe$_2$--MoSe$_2$ moiré heterostructure at filling $\nu=\frac{1}{2}$ and $\nu=1$, respectively.
Our work establishes a versatile and accessible platform for investigating exciton FCIs and opens a promising avenue towards the realization of non-Abelian anyons.

Importantly, our findings provide a specific guideline for the experimental realization of exciton FCIs in tWSe$_2$--MoSe$_2$.
Concretely, we predict that the exciton FCI remains stable across a wide range of experimental parameters and is particularly robust at $\theta \in [2^\circ, 2.5^\circ]$ and $\Delta \in [-5\ \text{meV}, 0\ \text{meV}]$.
The experimental know-how for fabricating moiré heterostructures and generating long-lived interlayer excitons is very mature~\cite{giuseppe_nature,jauregui_interlayer,tagarelli_natphot,correlated_exciton_insulator_science}.
The experimental challenges will likely lie on the detection of the topological order and perhaps on the control of the exciton density.
We note that, despite the exciton's neutral net charge, there are methods which enable exciton transport that could be utilized to detect these phases---the most promising are counterflow transport measurements~\cite{counterflow_macdonald,excitons_fqhe}, while other methods might involve spatial gradients in the out-of-plane electric field, dielectric environment, or strain profiles~\cite{kis_exciton_transistor,roberto_strain}. Future studies could also attempt to identify the optical fingerprints of these phases.
The experimental control of the density of interlayer excitons with microsecond lifetimes is steadily improving~\cite{tugen2025opticalinjection}.
We also note that introducing hBN spacers between electron and hole layers further enhances the lifetime and also allows for the direct injection of interlayer excitons via electrical gates~\cite{exciton_insulator_nat}.
Importantly, even if the exciton density is not accurately fixed to the relevant filling factor, the particles can, in principle, re-arrange themselves into FCI puddles (pinned by disorder) with the appropriate density, similar to the presumable origin of Hall resistance plateaus as a function of gate voltage in fermionic moiré FCIs.

Apart from prompting experimental efforts in tWSe$_2$--MoSe$_2$, our work motivates the search for exciton FCIs in other moiré structures. Furthermore, while we have only focused on two specific phases at $\nu=\frac{1}{2}$ and $\nu=1$, other phases remain unexplored. Besides other bosonic Laughlin and hierarchy states, the pursuit of additional non-Abelian phases such as the Moore--Read state at $\nu=\frac{1}{3}$ and the Read--Rezayi state~\cite{read_rezayi} at $\nu=\frac{3}{2}$ is particularly interesting---especially so for the latter as it hosts Fibonacci anyons which hold large promise for topological quantum computing.
Our work serves as a catalyst for the exploration of exotic physics including unprecedented phenomena unique to many-boson systems.

Finally, we note that many challenging aspects are yet to be addressed. For instance, investigating the competition between liquid- and crystal-like exciton FCIs as well as superfluids and supersolids would be crucial for the understanding of these phases~\cite{fisher_superfluid_insulator, Lu_bosonic_superflied_transition}.
Such phase transitions could be studied by adding hBN spacers between the electron layer and the hole bilayer to enhance the interaction strength (by reducing the electron--hole attraction).
In addition, while the spin-valley physics in the many-body exciton system is at the edge of current numerical capabilities, it is an important issue that must be tackled. Moreover, theoretical efforts exploring large filling factors might demand more advanced models considering deviations from the bosonic description of excitons~\cite{nonbosonic_excitons}. Last but not least, the phase diagram can be enriched by adding doping, which in combination with excitons constitutes a realization of mixed Bose--Fermi physics~\cite{imamoglu_feshbach,correlated_exciton_insulator_science,chernikov_THz_trions}. Our work shows that an idealized model can provide a faithful description of moiré exciton FCIs, suggesting a route to approach these formidable problems.

\begin{acknowledgements}
 We acknowledge useful discussions with Daniel Erkensten, Zhao Liu, Ahmed Abouelkomsan and Ataç İmamoğlu. This work was supported by the Swedish Research Council (VR, grant 2024-04567), the Wallenberg Scholars program of the Knut and Alice Wallenberg Foundation (2023.0256) and the G\"oran Gustafsson Foundation for Research in Natural Sciences and Medicine. 
 The computations were enabled by resources provided by the National Academic Infrastructure for Supercomputing in Sweden (NAISS), partially funded by the Swedish Research Council through grant agreement no. 2022-06725. In addition, we utilized the Sunrise HPC facility supported by the Technical Division of the Department of Physics, Stockholm University.
\end{acknowledgements}

\appendix

\section{Exciton--exciton interaction}
We take the method employed for deriving the single-particle moiré exciton Hamiltonian~\cite{dasSarma_topologicalExcitons} and extend it in order to obtain the Hamiltonian describing exciton--exciton interactions, which takes the form
\begin{equation}
    H_\text{x-x} = \frac{1}{2}\sum_{ll'\mathbf{q}}V_{ll'}(\mathbf{q}):\!\rho_l(\mathbf{q})\rho_{l'}(-\mathbf{q})\!: 
\end{equation}
Here, $\rho_l(\mathbf{q})=\sum_{\mathbf{Q}}X^\dagger_{l\mathbf{Q+q}}X^{\phantom{\dagger}}_{l\mathbf{Q}}$ is the exciton density operator, $::$ denotes normal order, and $V_{ll'}(\mathbf{q})$ is the exciton--exciton interaction potential.
In the first part of this work, we consider ideal contact interactions of the form $V_{ll'}(\mathbf{q})=V_0$. Later, we consider the realistic long-range interaction potential, which reads
\begin{align}
    V_{ll'}(\mathbf{q}) =& V^\text{e-e}_{ll'}(\mathbf{q})+V^\text{h-h}_{ll'}(\mathbf{q})-V^\text{e-h}_{ll'}(\mathbf{q})-[V^\text{e-h}_{l'l}(-\mathbf{q})]^*, \nonumber \\
    V^\text{e-e}_{ll'}(\mathbf{q}) =&  F_{ll'}(\alpha_\text{h}\mathbf{q},\alpha_\text{h}\mathbf{q})\frac{e_0^2}{2A\epsilon\epsilon_0|\mathbf{q}|}, \nonumber \\
    V^\text{h-h}_{ll'}(\mathbf{q}) =& F_{ll'}(\alpha_\text{e}\mathbf{q},\alpha_\text{e}\mathbf{q}) \frac{e_0^2}{2A\epsilon\epsilon_0|\mathbf{q}|} \mathrm{e}^{-|\mathbf{q}|d|l-l'|}, \nonumber \\
    V^\text{e-h}_{ll'}(\mathbf{q}) =& F_{ll'}(\alpha_\text{e}\mathbf{q},\alpha_\text{h}\mathbf{q}) \frac{e_0^2}{2A\epsilon\epsilon_0|\mathbf{q}|} \mathrm{e}^{-|\mathbf{q}|dl},
    \label{eq:interaction}
\end{align}
with $d=0.67$ nm and $\epsilon=3.8$~\cite{DasSarma_exciton_prl}.
The exciton--exciton interaction potential above contains individual contributions accounting for interactions between the charges (electrons and holes) in different excitons, i.e. it includes electron--electron, hole--hole, and electron--hole interaction channels. Here, $F_{ll'}(\mathbf{q}_1,\mathbf{q}_2)=\mathcal{F}_{ll,\mathbf{q}_1} \mathcal{F}_{l'l',-\mathbf{q}_2}$ has been introduced together with the exciton form factor $\mathcal{F}_{ll',\mathbf{q}}=\sum_\mathbf{k}\phi^*_{l\mathbf{k}}\phi_{l'\mathbf{k+q}}$, where $\phi_{l\mathbf{k}}$ is the wave function describing the relative electron--hole coordinate within a single exciton.
While terms arising from different exchange mechanisms should formally contribute to the interaction, here we have focused on the direct term which should dominate in the context of interlayer excitons~\cite{erkensten_interlayerExcitons,steinhoff_xxinteractions,kyriienko_excitons,samuel_bosonic}. A detailed derivation of the interaction potential can be found in the SM.

% \bibliography{reference}% Produces the bibliography via BibTeX.

\begin{thebibliography}{83}%
\makeatletter
\providecommand \@ifxundefined [1]{%
 \@ifx{#1\undefined}
}%
\providecommand \@ifnum [1]{%
 \ifnum #1\expandafter \@firstoftwo
 \else \expandafter \@secondoftwo
 \fi
}%
\providecommand \@ifx [1]{%
 \ifx #1\expandafter \@firstoftwo
 \else \expandafter \@secondoftwo
 \fi
}%
\providecommand \natexlab [1]{#1}%
\providecommand \enquote  [1]{``#1''}%
\providecommand \bibnamefont  [1]{#1}%
\providecommand \bibfnamefont [1]{#1}%
\providecommand \citenamefont [1]{#1}%
\providecommand \href@noop [0]{\@secondoftwo}%
\providecommand \href [0]{\begingroup \@sanitize@url \@href}%
\providecommand \@href[1]{\@@startlink{#1}\@@href}%
\providecommand \@@href[1]{\endgroup#1\@@endlink}%
\providecommand \@sanitize@url [0]{\catcode `\\12\catcode `\$12\catcode `\&12\catcode `\#12\catcode `\^12\catcode `\_12\catcode `\%12\relax}%
\providecommand \@@startlink[1]{}%
\providecommand \@@endlink[0]{}%
\providecommand \url  [0]{\begingroup\@sanitize@url \@url }%
\providecommand \@url [1]{\endgroup\@href {#1}{\urlprefix }}%
\providecommand \urlprefix  [0]{URL }%
\providecommand \Eprint [0]{\href }%
\providecommand \doibase [0]{https://doi.org/}%
\providecommand \selectlanguage [0]{\@gobble}%
\providecommand \bibinfo  [0]{\@secondoftwo}%
\providecommand \bibfield  [0]{\@secondoftwo}%
\providecommand \translation [1]{[#1]}%
\providecommand \BibitemOpen [0]{}%
\providecommand \bibitemStop [0]{}%
\providecommand \bibitemNoStop [0]{.\EOS\space}%
\providecommand \EOS [0]{\spacefactor3000\relax}%
\providecommand \BibitemShut  [1]{\csname bibitem#1\endcsname}%
\let\auto@bib@innerbib\@empty
%</preamble>
\bibitem [{\citenamefont {Ju}\ \emph {et~al.}(2024)\citenamefont {Ju}, \citenamefont {MacDonald}, \citenamefont {Mak}, \citenamefont {Sha},\ and\ \citenamefont {Xu}}]{fqahe_review}%
  \BibitemOpen
  \bibfield  {author} {\bibinfo {author} {\bibfnamefont {L.}~\bibnamefont {Ju}}, \bibinfo {author} {\bibfnamefont {A.~H.}\ \bibnamefont {MacDonald}}, \bibinfo {author} {\bibfnamefont {K.~F.}\ \bibnamefont {Mak}}, \bibinfo {author} {\bibfnamefont {J.}~\bibnamefont {Sha}},\ and\ \bibinfo {author} {\bibfnamefont {X.}~\bibnamefont {Xu}},\ }\bibfield  {title} {\bibinfo {title} {The fractional quantum anomalous {Hall} effect},\ }\href {https://doi.org/10.1038/s41578-024-00694-x} {\bibfield  {journal} {\bibinfo  {journal} {Nature Reviews Materials}\ ,\ \bibinfo {pages} {455--459}} (\bibinfo {year} {2024})}\BibitemShut {NoStop}%
\bibitem [{\citenamefont {Parameswaran}\ \emph {et~al.}(2013)\citenamefont {Parameswaran}, \citenamefont {Roy},\ and\ \citenamefont {Sondhi}}]{PARAMESWARAN2013816}%
  \BibitemOpen
  \bibfield  {author} {\bibinfo {author} {\bibfnamefont {S.~A.}\ \bibnamefont {Parameswaran}}, \bibinfo {author} {\bibfnamefont {R.}~\bibnamefont {Roy}},\ and\ \bibinfo {author} {\bibfnamefont {S.~L.}\ \bibnamefont {Sondhi}},\ }\bibfield  {title} {\bibinfo {title} {Fractional quantum {Hall} physics in topological flat bands},\ }\href {https://doi.org/https://doi.org/10.1016/j.crhy.2013.04.003} {\bibfield  {journal} {\bibinfo  {journal} {Comptes Rendus Physique}\ }\textbf {\bibinfo {volume} {14}},\ \bibinfo {pages} {816--839} (\bibinfo {year} {2013})},\ \bibinfo {note} {topological insulators / Isolants topologiques}\BibitemShut {NoStop}%
\bibitem [{\citenamefont {Liu}\ and\ \citenamefont {Bergholtz}(2024)}]{Zhao_Liu_Review}%
  \BibitemOpen
  \bibfield  {author} {\bibinfo {author} {\bibfnamefont {Z.}~\bibnamefont {Liu}}\ and\ \bibinfo {author} {\bibfnamefont {E.~J.}\ \bibnamefont {Bergholtz}},\ }\bibfield  {title} {\bibinfo {title} {Recent developments in fractional {Chern} insulators},\ }in\ \href {https://doi.org/https://doi.org/10.1016/B978-0-323-90800-9.00136-0} {\emph {\bibinfo {booktitle} {Encyclopedia of Condensed Matter Physics (Second Edition)}}},\ \bibinfo {editor} {edited by\ \bibinfo {editor} {\bibfnamefont {T.}~\bibnamefont {Chakraborty}}}\ (\bibinfo  {publisher} {Academic Press},\ \bibinfo {address} {Oxford},\ \bibinfo {year} {2024})\ \bibinfo {edition} {second edition}\ ed.,\ pp.\ \bibinfo {pages} {515--538}\BibitemShut {NoStop}%
\bibitem [{\citenamefont {Spanton}\ \emph {et~al.}(2018)\citenamefont {Spanton}, \citenamefont {Zibrov}, \citenamefont {Zhou}, \citenamefont {Taniguchi}, \citenamefont {Watanabe}, \citenamefont {Zaletel},\ and\ \citenamefont {Young}}]{spanton2018observation}%
  \BibitemOpen
  \bibfield  {author} {\bibinfo {author} {\bibfnamefont {E.~M.}\ \bibnamefont {Spanton}}, \bibinfo {author} {\bibfnamefont {A.~A.}\ \bibnamefont {Zibrov}}, \bibinfo {author} {\bibfnamefont {H.}~\bibnamefont {Zhou}}, \bibinfo {author} {\bibfnamefont {T.}~\bibnamefont {Taniguchi}}, \bibinfo {author} {\bibfnamefont {K.}~\bibnamefont {Watanabe}}, \bibinfo {author} {\bibfnamefont {M.~P.}\ \bibnamefont {Zaletel}},\ and\ \bibinfo {author} {\bibfnamefont {A.~F.}\ \bibnamefont {Young}},\ }\bibfield  {title} {\bibinfo {title} {Observation of fractional {Chern} insulators in a van der {Waals} heterostructure},\ }\href {https://www.science.org/doi/10.1126/science.aan8458} {\bibfield  {journal} {\bibinfo  {journal} {Science}\ }\textbf {\bibinfo {volume} {360}},\ \bibinfo {pages} {62--66} (\bibinfo {year} {2018})}\BibitemShut {NoStop}%
\bibitem [{\citenamefont {Xie}\ \emph {et~al.}(2021)\citenamefont {Xie}, \citenamefont {Pierce}, \citenamefont {Park}, \citenamefont {Parker}, \citenamefont {Khalaf}, \citenamefont {Ledwith}, \citenamefont {Cao}, \citenamefont {Lee}, \citenamefont {Chen}, \citenamefont {Forrester}, \citenamefont {Watanabe}, \citenamefont {Taniguchi}, \citenamefont {Vishwanath}, \citenamefont {Jarillo-Herrero},\ and\ \citenamefont {Yacoby}}]{xie2021fractional}%
  \BibitemOpen
  \bibfield  {author} {\bibinfo {author} {\bibfnamefont {Y.}~\bibnamefont {Xie}}, \bibinfo {author} {\bibfnamefont {A.~T.}\ \bibnamefont {Pierce}}, \bibinfo {author} {\bibfnamefont {J.~M.}\ \bibnamefont {Park}}, \bibinfo {author} {\bibfnamefont {D.~E.}\ \bibnamefont {Parker}}, \bibinfo {author} {\bibfnamefont {E.}~\bibnamefont {Khalaf}}, \bibinfo {author} {\bibfnamefont {P.}~\bibnamefont {Ledwith}}, \bibinfo {author} {\bibfnamefont {Y.}~\bibnamefont {Cao}}, \bibinfo {author} {\bibfnamefont {S.~H.}\ \bibnamefont {Lee}}, \bibinfo {author} {\bibfnamefont {S.}~\bibnamefont {Chen}}, \bibinfo {author} {\bibfnamefont {P.~R.}\ \bibnamefont {Forrester}}, \bibinfo {author} {\bibfnamefont {K.}~\bibnamefont {Watanabe}}, \bibinfo {author} {\bibfnamefont {T.}~\bibnamefont {Taniguchi}}, \bibinfo {author} {\bibfnamefont {A.}~\bibnamefont {Vishwanath}}, \bibinfo {author} {\bibfnamefont {P.}~\bibnamefont {Jarillo-Herrero}},\ and\ \bibinfo {author} {\bibfnamefont {A.}~\bibnamefont {Yacoby}},\ }\bibfield  {title} {\bibinfo
  {title} {Fractional {Chern} insulators in magic-angle twisted bilayer graphene},\ }\href {https://doi.org/https://doi.org/10.1038/s41586-021-04002-3} {\bibfield  {journal} {\bibinfo  {journal} {Nature}\ }\textbf {\bibinfo {volume} {600}},\ \bibinfo {pages} {439--443} (\bibinfo {year} {2021})}\BibitemShut {NoStop}%
\bibitem [{\citenamefont {Abouelkomsan}\ \emph {et~al.}(2020)\citenamefont {Abouelkomsan}, \citenamefont {Liu},\ and\ \citenamefont {Bergholtz}}]{ahmed_fci}%
  \BibitemOpen
  \bibfield  {author} {\bibinfo {author} {\bibfnamefont {A.}~\bibnamefont {Abouelkomsan}}, \bibinfo {author} {\bibfnamefont {Z.}~\bibnamefont {Liu}},\ and\ \bibinfo {author} {\bibfnamefont {E.~J.}\ \bibnamefont {Bergholtz}},\ }\bibfield  {title} {\bibinfo {title} {Particle-hole duality, emergent {Fermi} liquids, and fractional {Chern} insulators in moir\'e flatbands},\ }\href {https://doi.org/10.1103/PhysRevLett.124.106803} {\bibfield  {journal} {\bibinfo  {journal} {Phys. Rev. Lett.}\ }\textbf {\bibinfo {volume} {124}},\ \bibinfo {pages} {106803} (\bibinfo {year} {2020})}\BibitemShut {NoStop}%
\bibitem [{\citenamefont {Repellin}\ and\ \citenamefont {Senthil}(2020)}]{repellinChernBandsTwisted2020}%
  \BibitemOpen
  \bibfield  {author} {\bibinfo {author} {\bibfnamefont {C.}~\bibnamefont {Repellin}}\ and\ \bibinfo {author} {\bibfnamefont {T.}~\bibnamefont {Senthil}},\ }\bibfield  {title} {\bibinfo {title} {{Chern} bands of twisted bilayer graphene: {Fractional Chern} insulators and spin phase transition},\ }\href {https://doi.org/10.1103/PhysRevResearch.2.023238} {\bibfield  {journal} {\bibinfo  {journal} {Physical Review Research}\ }\textbf {\bibinfo {volume} {2}},\ \bibinfo {pages} {023238} (\bibinfo {year} {2020})}\BibitemShut {NoStop}%
\bibitem [{\citenamefont {Ledwith}\ \emph {et~al.}(2020)\citenamefont {Ledwith}, \citenamefont {Tarnopolsky}, \citenamefont {Khalaf},\ and\ \citenamefont {Vishwanath}}]{Ledwith_analytical_approach_FCI}%
  \BibitemOpen
  \bibfield  {author} {\bibinfo {author} {\bibfnamefont {P.~J.}\ \bibnamefont {Ledwith}}, \bibinfo {author} {\bibfnamefont {G.}~\bibnamefont {Tarnopolsky}}, \bibinfo {author} {\bibfnamefont {E.}~\bibnamefont {Khalaf}},\ and\ \bibinfo {author} {\bibfnamefont {A.}~\bibnamefont {Vishwanath}},\ }\bibfield  {title} {\bibinfo {title} {Fractional {Chern} insulator states in twisted bilayer graphene: {An} analytical approach},\ }\href {https://doi.org/10.1103/PhysRevResearch.2.023237} {\bibfield  {journal} {\bibinfo  {journal} {Phys. Rev. Res.}\ }\textbf {\bibinfo {volume} {2}},\ \bibinfo {pages} {023237} (\bibinfo {year} {2020})}\BibitemShut {NoStop}%
\bibitem [{\citenamefont {Park}\ \emph {et~al.}(2023)\citenamefont {Park}, \citenamefont {Cai}, \citenamefont {Anderson}, \citenamefont {Zhang}, \citenamefont {Zhu}, \citenamefont {Liu}, \citenamefont {Wang}, \citenamefont {Holtzmann}, \citenamefont {Hu}, \citenamefont {Liu}, \citenamefont {Taniguchi}, \citenamefont {Watanabe}, \citenamefont {Chu}, \citenamefont {Cao}, \citenamefont {Fu}, \citenamefont {Yao}, \citenamefont {Chang}, \citenamefont {Cobden}, \citenamefont {Xiao},\ and\ \citenamefont {Xu}}]{FCI_MoTe2_3}%
  \BibitemOpen
  \bibfield  {author} {\bibinfo {author} {\bibfnamefont {H.}~\bibnamefont {Park}}, \bibinfo {author} {\bibfnamefont {J.}~\bibnamefont {Cai}}, \bibinfo {author} {\bibfnamefont {E.}~\bibnamefont {Anderson}}, \bibinfo {author} {\bibfnamefont {Y.}~\bibnamefont {Zhang}}, \bibinfo {author} {\bibfnamefont {J.}~\bibnamefont {Zhu}}, \bibinfo {author} {\bibfnamefont {X.}~\bibnamefont {Liu}}, \bibinfo {author} {\bibfnamefont {C.}~\bibnamefont {Wang}}, \bibinfo {author} {\bibfnamefont {W.}~\bibnamefont {Holtzmann}}, \bibinfo {author} {\bibfnamefont {C.}~\bibnamefont {Hu}}, \bibinfo {author} {\bibfnamefont {Z.}~\bibnamefont {Liu}}, \bibinfo {author} {\bibfnamefont {T.}~\bibnamefont {Taniguchi}}, \bibinfo {author} {\bibfnamefont {K.}~\bibnamefont {Watanabe}}, \bibinfo {author} {\bibfnamefont {J.-H.}\ \bibnamefont {Chu}}, \bibinfo {author} {\bibfnamefont {T.}~\bibnamefont {Cao}}, \bibinfo {author} {\bibfnamefont {L.}~\bibnamefont {Fu}}, \bibinfo {author} {\bibfnamefont {W.}~\bibnamefont {Yao}}, \bibinfo {author}
  {\bibfnamefont {C.-Z.}\ \bibnamefont {Chang}}, \bibinfo {author} {\bibfnamefont {D.}~\bibnamefont {Cobden}}, \bibinfo {author} {\bibfnamefont {D.}~\bibnamefont {Xiao}},\ and\ \bibinfo {author} {\bibfnamefont {X.}~\bibnamefont {Xu}},\ }\bibfield  {title} {\bibinfo {title} {Observation of fractionally quantized anomalous {Hall} effect},\ }\href {https://doi.org/10.1038/s41586-023-06536-0} {\bibfield  {journal} {\bibinfo  {journal} {Nature}\ }\textbf {\bibinfo {volume} {622}},\ \bibinfo {pages} {74--79} (\bibinfo {year} {2023})}\BibitemShut {NoStop}%
\bibitem [{\citenamefont {Zeng}\ \emph {et~al.}(2023)\citenamefont {Zeng}, \citenamefont {Xia}, \citenamefont {Kang}, \citenamefont {Zhu}, \citenamefont {Kn{\"u}ppel}, \citenamefont {Vaswani}, \citenamefont {Watanabe}, \citenamefont {Taniguchi}, \citenamefont {Mak},\ and\ \citenamefont {Shan}}]{FCI_MoTe2_2}%
  \BibitemOpen
  \bibfield  {author} {\bibinfo {author} {\bibfnamefont {Y.}~\bibnamefont {Zeng}}, \bibinfo {author} {\bibfnamefont {Z.}~\bibnamefont {Xia}}, \bibinfo {author} {\bibfnamefont {K.}~\bibnamefont {Kang}}, \bibinfo {author} {\bibfnamefont {J.}~\bibnamefont {Zhu}}, \bibinfo {author} {\bibfnamefont {P.}~\bibnamefont {Kn{\"u}ppel}}, \bibinfo {author} {\bibfnamefont {C.}~\bibnamefont {Vaswani}}, \bibinfo {author} {\bibfnamefont {K.}~\bibnamefont {Watanabe}}, \bibinfo {author} {\bibfnamefont {T.}~\bibnamefont {Taniguchi}}, \bibinfo {author} {\bibfnamefont {K.~F.}\ \bibnamefont {Mak}},\ and\ \bibinfo {author} {\bibfnamefont {J.}~\bibnamefont {Shan}},\ }\bibfield  {title} {\bibinfo {title} {Thermodynamic evidence of fractional {Chern} insulator in moir{\'e} {MoTe}$_2$},\ }\href {https://doi.org/10.1038/s41586-023-06452-3} {\bibfield  {journal} {\bibinfo  {journal} {Nature}\ }\textbf {\bibinfo {volume} {622}},\ \bibinfo {pages} {69--73} (\bibinfo {year} {2023})}\BibitemShut {NoStop}%
\bibitem [{\citenamefont {Xu}\ \emph {et~al.}(2023)\citenamefont {Xu}, \citenamefont {Sun}, \citenamefont {Jia}, \citenamefont {Liu}, \citenamefont {Xu}, \citenamefont {Li}, \citenamefont {Gu}, \citenamefont {Watanabe}, \citenamefont {Taniguchi}, \citenamefont {Tong}, \citenamefont {Jia}, \citenamefont {Shi}, \citenamefont {Jiang}, \citenamefont {Zhang}, \citenamefont {Liu},\ and\ \citenamefont {Li}}]{PhysRevX.13.031037}%
  \BibitemOpen
  \bibfield  {author} {\bibinfo {author} {\bibfnamefont {F.}~\bibnamefont {Xu}}, \bibinfo {author} {\bibfnamefont {Z.}~\bibnamefont {Sun}}, \bibinfo {author} {\bibfnamefont {T.}~\bibnamefont {Jia}}, \bibinfo {author} {\bibfnamefont {C.}~\bibnamefont {Liu}}, \bibinfo {author} {\bibfnamefont {C.}~\bibnamefont {Xu}}, \bibinfo {author} {\bibfnamefont {C.}~\bibnamefont {Li}}, \bibinfo {author} {\bibfnamefont {Y.}~\bibnamefont {Gu}}, \bibinfo {author} {\bibfnamefont {K.}~\bibnamefont {Watanabe}}, \bibinfo {author} {\bibfnamefont {T.}~\bibnamefont {Taniguchi}}, \bibinfo {author} {\bibfnamefont {B.}~\bibnamefont {Tong}}, \bibinfo {author} {\bibfnamefont {J.}~\bibnamefont {Jia}}, \bibinfo {author} {\bibfnamefont {Z.}~\bibnamefont {Shi}}, \bibinfo {author} {\bibfnamefont {S.}~\bibnamefont {Jiang}}, \bibinfo {author} {\bibfnamefont {Y.}~\bibnamefont {Zhang}}, \bibinfo {author} {\bibfnamefont {X.}~\bibnamefont {Liu}},\ and\ \bibinfo {author} {\bibfnamefont {T.}~\bibnamefont {Li}},\ }\bibfield  {title} {\bibinfo {title}
  {Observation of integer and fractional quantum anomalous {Hall} effects in twisted bilayer {MoTe}$_{2}$},\ }\href {https://doi.org/10.1103/PhysRevX.13.031037} {\bibfield  {journal} {\bibinfo  {journal} {Phys. Rev. X}\ }\textbf {\bibinfo {volume} {13}},\ \bibinfo {pages} {031037} (\bibinfo {year} {2023})}\BibitemShut {NoStop}%
\bibitem [{\citenamefont {Lu}\ \emph {et~al.}(2024)\citenamefont {Lu}, \citenamefont {Han}, \citenamefont {Yao}, \citenamefont {Reddy}, \citenamefont {Yang}, \citenamefont {Seo}, \citenamefont {Watanabe}, \citenamefont {Taniguchi}, \citenamefont {Fu},\ and\ \citenamefont {Ju}}]{lu2024fractional}%
  \BibitemOpen
  \bibfield  {author} {\bibinfo {author} {\bibfnamefont {Z.}~\bibnamefont {Lu}}, \bibinfo {author} {\bibfnamefont {T.}~\bibnamefont {Han}}, \bibinfo {author} {\bibfnamefont {Y.}~\bibnamefont {Yao}}, \bibinfo {author} {\bibfnamefont {A.~P.}\ \bibnamefont {Reddy}}, \bibinfo {author} {\bibfnamefont {J.}~\bibnamefont {Yang}}, \bibinfo {author} {\bibfnamefont {J.}~\bibnamefont {Seo}}, \bibinfo {author} {\bibfnamefont {K.}~\bibnamefont {Watanabe}}, \bibinfo {author} {\bibfnamefont {T.}~\bibnamefont {Taniguchi}}, \bibinfo {author} {\bibfnamefont {L.}~\bibnamefont {Fu}},\ and\ \bibinfo {author} {\bibfnamefont {L.}~\bibnamefont {Ju}},\ }\bibfield  {title} {\bibinfo {title} {Fractional quantum anomalous {Hall} effect in multilayer graphene},\ }\href {https://doi.org/10.1038/s41586-023-07010-7} {\bibfield  {journal} {\bibinfo  {journal} {Nature}\ }\textbf {\bibinfo {volume} {626}},\ \bibinfo {pages} {759--764} (\bibinfo {year} {2024})}\BibitemShut {NoStop}%
\bibitem [{\citenamefont {Reddy}\ \emph {et~al.}(2024)\citenamefont {Reddy}, \citenamefont {Paul}, \citenamefont {Abouelkomsan},\ and\ \citenamefont {Fu}}]{mr_Aidan}%
  \BibitemOpen
  \bibfield  {author} {\bibinfo {author} {\bibfnamefont {A.~P.}\ \bibnamefont {Reddy}}, \bibinfo {author} {\bibfnamefont {N.}~\bibnamefont {Paul}}, \bibinfo {author} {\bibfnamefont {A.}~\bibnamefont {Abouelkomsan}},\ and\ \bibinfo {author} {\bibfnamefont {L.}~\bibnamefont {Fu}},\ }\bibfield  {title} {\bibinfo {title} {Non-{Abelian} fractionalization in topological minibands},\ }\href {https://doi.org/10.1103/PhysRevLett.133.166503} {\bibfield  {journal} {\bibinfo  {journal} {Phys. Rev. Lett.}\ }\textbf {\bibinfo {volume} {133}},\ \bibinfo {pages} {166503} (\bibinfo {year} {2024})}\BibitemShut {NoStop}%
\bibitem [{\citenamefont {Ahn}\ \emph {et~al.}(2024)\citenamefont {Ahn}, \citenamefont {Lee}, \citenamefont {Yananose}, \citenamefont {Kim},\ and\ \citenamefont {Cho}}]{mr_Ahn}%
  \BibitemOpen
  \bibfield  {author} {\bibinfo {author} {\bibfnamefont {C.-E.}\ \bibnamefont {Ahn}}, \bibinfo {author} {\bibfnamefont {W.}~\bibnamefont {Lee}}, \bibinfo {author} {\bibfnamefont {K.}~\bibnamefont {Yananose}}, \bibinfo {author} {\bibfnamefont {Y.}~\bibnamefont {Kim}},\ and\ \bibinfo {author} {\bibfnamefont {G.~Y.}\ \bibnamefont {Cho}},\ }\bibfield  {title} {\bibinfo {title} {Non-{Abelian} fractional quantum anomalous {Hall} states and first {Landau} level physics of the second moir\'e band of twisted bilayer {MoTe}$_{2}$},\ }\href {https://doi.org/10.1103/PhysRevB.110.L161109} {\bibfield  {journal} {\bibinfo  {journal} {Phys. Rev. B}\ }\textbf {\bibinfo {volume} {110}},\ \bibinfo {pages} {L161109} (\bibinfo {year} {2024})}\BibitemShut {NoStop}%
\bibitem [{\citenamefont {Chen}\ \emph {et~al.}(2025)\citenamefont {Chen}, \citenamefont {Luo}, \citenamefont {Zhu},\ and\ \citenamefont {Sheng}}]{mr_Donna}%
  \BibitemOpen
  \bibfield  {author} {\bibinfo {author} {\bibfnamefont {F.}~\bibnamefont {Chen}}, \bibinfo {author} {\bibfnamefont {W.-W.}\ \bibnamefont {Luo}}, \bibinfo {author} {\bibfnamefont {W.}~\bibnamefont {Zhu}},\ and\ \bibinfo {author} {\bibfnamefont {D.~N.}\ \bibnamefont {Sheng}},\ }\bibfield  {title} {\bibinfo {title} {Robust non-{Abelian} even-denominator fractional {Chern} insulator in twisted bilayer {MoTe}$_2$},\ }\href {https://doi.org/10.1038/s41467-025-57326-3} {\bibfield  {journal} {\bibinfo  {journal} {Nature Communications}\ }\textbf {\bibinfo {volume} {16}},\ \bibinfo {pages} {2115} (\bibinfo {year} {2025})}\BibitemShut {NoStop}%
\bibitem [{\citenamefont {Liu}\ \emph {et~al.}(2025{\natexlab{a}})\citenamefont {Liu}, \citenamefont {Liu},\ and\ \citenamefont {Bergholtz}}]{mr_Hui}%
  \BibitemOpen
  \bibfield  {author} {\bibinfo {author} {\bibfnamefont {H.}~\bibnamefont {Liu}}, \bibinfo {author} {\bibfnamefont {Z.}~\bibnamefont {Liu}},\ and\ \bibinfo {author} {\bibfnamefont {E.~J.}\ \bibnamefont {Bergholtz}},\ }\bibfield  {title} {\bibinfo {title} {{Non-Abelian Fractional Chern Insulators and Competing States in Flat Moir\'e Bands}},\ }\href {https://link.aps.org/doi/10.1103/43nq-ntqm} {\bibfield  {journal} {\bibinfo  {journal} {Phys. Rev. Lett.}\ }\textbf {\bibinfo {volume} {135}},\ \bibinfo {pages} {106604} (\bibinfo {year} {2025}{\natexlab{a}})}\BibitemShut {NoStop}%
\bibitem [{\citenamefont {Xu}\ \emph {et~al.}(2025)\citenamefont {Xu}, \citenamefont {Mao}, \citenamefont {Zeng},\ and\ \citenamefont {Zhang}}]{mr_Xu}%
  \BibitemOpen
  \bibfield  {author} {\bibinfo {author} {\bibfnamefont {C.}~\bibnamefont {Xu}}, \bibinfo {author} {\bibfnamefont {N.}~\bibnamefont {Mao}}, \bibinfo {author} {\bibfnamefont {T.}~\bibnamefont {Zeng}},\ and\ \bibinfo {author} {\bibfnamefont {Y.}~\bibnamefont {Zhang}},\ }\bibfield  {title} {\bibinfo {title} {Multiple {Chern} bands in twisted {MoTe}$_2$ and possible {Non-Abelian} states},\ }\href {https://doi.org/10.1103/PhysRevLett.134.066601} {\bibfield  {journal} {\bibinfo  {journal} {Phys. Rev. Lett.}\ }\textbf {\bibinfo {volume} {134}},\ \bibinfo {pages} {066601} (\bibinfo {year} {2025})}\BibitemShut {NoStop}%
\bibitem [{\citenamefont {Wang}\ \emph {et~al.}(2025)\citenamefont {Wang}, \citenamefont {Zhang}, \citenamefont {Liu}, \citenamefont {Wang}, \citenamefont {Cao},\ and\ \citenamefont {Xiao}}]{mr_Wang}%
  \BibitemOpen
  \bibfield  {author} {\bibinfo {author} {\bibfnamefont {C.}~\bibnamefont {Wang}}, \bibinfo {author} {\bibfnamefont {X.-W.}\ \bibnamefont {Zhang}}, \bibinfo {author} {\bibfnamefont {X.}~\bibnamefont {Liu}}, \bibinfo {author} {\bibfnamefont {J.}~\bibnamefont {Wang}}, \bibinfo {author} {\bibfnamefont {T.}~\bibnamefont {Cao}},\ and\ \bibinfo {author} {\bibfnamefont {D.}~\bibnamefont {Xiao}},\ }\bibfield  {title} {\bibinfo {title} {Higher {Landau}-level analogs and signatures of non-{Abelian} states in twisted bilayer {MoTe}$_2$},\ }\href {https://doi.org/10.1103/PhysRevLett.134.076503} {\bibfield  {journal} {\bibinfo  {journal} {Phys. Rev. Lett.}\ }\textbf {\bibinfo {volume} {134}},\ \bibinfo {pages} {076503} (\bibinfo {year} {2025})}\BibitemShut {NoStop}%
\bibitem [{\citenamefont {Liu}\ \emph {et~al.}(2025{\natexlab{b}})\citenamefont {Liu}, \citenamefont {Perea-Causin},\ and\ \citenamefont {Bergholtz}}]{our_parafermion}%
  \BibitemOpen
  \bibfield  {author} {\bibinfo {author} {\bibfnamefont {H.}~\bibnamefont {Liu}}, \bibinfo {author} {\bibfnamefont {R.}~\bibnamefont {Perea-Causin}},\ and\ \bibinfo {author} {\bibfnamefont {E.~J.}\ \bibnamefont {Bergholtz}},\ }\bibfield  {title} {\bibinfo {title} {Parafermions in moiré minibands},\ }\href {https://doi.org/10.1038/s41467-025-57035-x} {\bibfield  {journal} {\bibinfo  {journal} {Nature Communications}\ }\textbf {\bibinfo {volume} {16}},\ \bibinfo {pages} {1770} (\bibinfo {year} {2025}{\natexlab{b}})}\BibitemShut {NoStop}%
\bibitem [{\citenamefont {Nayak}\ \emph {et~al.}(2008)\citenamefont {Nayak}, \citenamefont {Simon}, \citenamefont {Stern}, \citenamefont {Freedman},\ and\ \citenamefont {Das~Sarma}}]{RevModPhys.80.1083}%
  \BibitemOpen
  \bibfield  {author} {\bibinfo {author} {\bibfnamefont {C.}~\bibnamefont {Nayak}}, \bibinfo {author} {\bibfnamefont {S.~H.}\ \bibnamefont {Simon}}, \bibinfo {author} {\bibfnamefont {A.}~\bibnamefont {Stern}}, \bibinfo {author} {\bibfnamefont {M.}~\bibnamefont {Freedman}},\ and\ \bibinfo {author} {\bibfnamefont {S.}~\bibnamefont {Das~Sarma}},\ }\bibfield  {title} {\bibinfo {title} {Non-{Abelian} anyons and topological quantum computation},\ }\href {https://doi.org/10.1103/RevModPhys.80.1083} {\bibfield  {journal} {\bibinfo  {journal} {Rev. Mod. Phys.}\ }\textbf {\bibinfo {volume} {80}},\ \bibinfo {pages} {1083--1159} (\bibinfo {year} {2008})}\BibitemShut {NoStop}%
\bibitem [{\citenamefont {Wang}\ \emph {et~al.}(2018{\natexlab{a}})\citenamefont {Wang}, \citenamefont {Chernikov}, \citenamefont {Glazov}, \citenamefont {Heinz}, \citenamefont {Marie}, \citenamefont {Amand},\ and\ \citenamefont {Urbaszek}}]{excitons_colloquium}%
  \BibitemOpen
  \bibfield  {author} {\bibinfo {author} {\bibfnamefont {G.}~\bibnamefont {Wang}}, \bibinfo {author} {\bibfnamefont {A.}~\bibnamefont {Chernikov}}, \bibinfo {author} {\bibfnamefont {M.~M.}\ \bibnamefont {Glazov}}, \bibinfo {author} {\bibfnamefont {T.~F.}\ \bibnamefont {Heinz}}, \bibinfo {author} {\bibfnamefont {X.}~\bibnamefont {Marie}}, \bibinfo {author} {\bibfnamefont {T.}~\bibnamefont {Amand}},\ and\ \bibinfo {author} {\bibfnamefont {B.}~\bibnamefont {Urbaszek}},\ }\bibfield  {title} {\bibinfo {title} {Colloquium: {Excitons} in atomically thin transition metal dichalcogenides},\ }\href {https://doi.org/10.1103/RevModPhys.90.021001} {\bibfield  {journal} {\bibinfo  {journal} {Rev. Mod. Phys.}\ }\textbf {\bibinfo {volume} {90}},\ \bibinfo {pages} {021001} (\bibinfo {year} {2018}{\natexlab{a}})}\BibitemShut {NoStop}%
\bibitem [{\citenamefont {Mak}\ and\ \citenamefont {Sha}(2022)}]{semiconductor_moire_review}%
  \BibitemOpen
  \bibfield  {author} {\bibinfo {author} {\bibfnamefont {K.~F.}\ \bibnamefont {Mak}}\ and\ \bibinfo {author} {\bibfnamefont {J.}~\bibnamefont {Sha}},\ }\bibfield  {title} {\bibinfo {title} {Semiconductor moiré materials},\ }\href {https://doi.org/10.1038/s41565-022-01165-6} {\bibfield  {journal} {\bibinfo  {journal} {Nature Nanotechnology}\ ,\ \bibinfo {pages} {686--695}} (\bibinfo {year} {2022})}\BibitemShut {NoStop}%
\bibitem [{\citenamefont {Regan}\ \emph {et~al.}(2022)\citenamefont {Regan}, \citenamefont {Wang}, \citenamefont {Paik}, \citenamefont {Zeng}, \citenamefont {Zhang}, \citenamefont {Zhu}, \citenamefont {MacDonald}, \citenamefont {Deng},\ and\ \citenamefont {Wang}}]{exciton_TMDbilayers_review}%
  \BibitemOpen
  \bibfield  {author} {\bibinfo {author} {\bibfnamefont {E.~C.}\ \bibnamefont {Regan}}, \bibinfo {author} {\bibfnamefont {D.}~\bibnamefont {Wang}}, \bibinfo {author} {\bibfnamefont {E.~Y.}\ \bibnamefont {Paik}}, \bibinfo {author} {\bibfnamefont {Y.}~\bibnamefont {Zeng}}, \bibinfo {author} {\bibfnamefont {L.}~\bibnamefont {Zhang}}, \bibinfo {author} {\bibfnamefont {J.}~\bibnamefont {Zhu}}, \bibinfo {author} {\bibfnamefont {A.~H.}\ \bibnamefont {MacDonald}}, \bibinfo {author} {\bibfnamefont {H.}~\bibnamefont {Deng}},\ and\ \bibinfo {author} {\bibfnamefont {F.}~\bibnamefont {Wang}},\ }\bibfield  {title} {\bibinfo {title} {Emerging exciton physics in transition metal dichalcogenide heterobilayers},\ }\href {https://doi.org/10.1038/s41578-022-00440-1} {\bibfield  {journal} {\bibinfo  {journal} {Nature Reviews Materials}\ ,\ \bibinfo {pages} {778--795}} (\bibinfo {year} {2022})}\BibitemShut {NoStop}%
\bibitem [{\citenamefont {Jauregui}\ \emph {et~al.}(2019)\citenamefont {Jauregui}, \citenamefont {Joe}, \citenamefont {Pistunova}, \citenamefont {Wild}, \citenamefont {High}, \citenamefont {Zhou}, \citenamefont {Scuri}, \citenamefont {Greve}, \citenamefont {Sushko}, \citenamefont {Yu}, \citenamefont {Taniguchi}, \citenamefont {Watanabe}, \citenamefont {Needleman}, \citenamefont {Lukin}, \citenamefont {Park},\ and\ \citenamefont {Kim}}]{jauregui_interlayer}%
  \BibitemOpen
  \bibfield  {author} {\bibinfo {author} {\bibfnamefont {L.~A.}\ \bibnamefont {Jauregui}}, \bibinfo {author} {\bibfnamefont {A.~Y.}\ \bibnamefont {Joe}}, \bibinfo {author} {\bibfnamefont {K.}~\bibnamefont {Pistunova}}, \bibinfo {author} {\bibfnamefont {D.~S.}\ \bibnamefont {Wild}}, \bibinfo {author} {\bibfnamefont {A.~A.}\ \bibnamefont {High}}, \bibinfo {author} {\bibfnamefont {Y.}~\bibnamefont {Zhou}}, \bibinfo {author} {\bibfnamefont {G.}~\bibnamefont {Scuri}}, \bibinfo {author} {\bibfnamefont {K.~D.}\ \bibnamefont {Greve}}, \bibinfo {author} {\bibfnamefont {A.}~\bibnamefont {Sushko}}, \bibinfo {author} {\bibfnamefont {C.-H.}\ \bibnamefont {Yu}}, \bibinfo {author} {\bibfnamefont {T.}~\bibnamefont {Taniguchi}}, \bibinfo {author} {\bibfnamefont {K.}~\bibnamefont {Watanabe}}, \bibinfo {author} {\bibfnamefont {D.~J.}\ \bibnamefont {Needleman}}, \bibinfo {author} {\bibfnamefont {M.~D.}\ \bibnamefont {Lukin}}, \bibinfo {author} {\bibfnamefont {H.}~\bibnamefont {Park}},\ and\ \bibinfo {author} {\bibfnamefont
  {P.}~\bibnamefont {Kim}},\ }\bibfield  {title} {\bibinfo {title} {Electrical control of interlayer exciton dynamics in atomically thin heterostructures},\ }\href {https://doi.org/10.1126/science.aaw4194} {\bibfield  {journal} {\bibinfo  {journal} {Science}\ }\textbf {\bibinfo {volume} {366}},\ \bibinfo {pages} {870--875} (\bibinfo {year} {2019})}\BibitemShut {NoStop}%
\bibitem [{\citenamefont {Choi}\ \emph {et~al.}(2021)\citenamefont {Choi}, \citenamefont {Florian}, \citenamefont {Steinhoff}, \citenamefont {Erben}, \citenamefont {Tran}, \citenamefont {Kim}, \citenamefont {Sun}, \citenamefont {Quan}, \citenamefont {Claassen}, \citenamefont {Majumder}, \citenamefont {Hollingsworth}, \citenamefont {Taniguchi}, \citenamefont {Watanabe}, \citenamefont {Ueno}, \citenamefont {Singh}, \citenamefont {Moody}, \citenamefont {Jahnke},\ and\ \citenamefont {Li}}]{elaine_lifetime}%
  \BibitemOpen
  \bibfield  {author} {\bibinfo {author} {\bibfnamefont {J.}~\bibnamefont {Choi}}, \bibinfo {author} {\bibfnamefont {M.}~\bibnamefont {Florian}}, \bibinfo {author} {\bibfnamefont {A.}~\bibnamefont {Steinhoff}}, \bibinfo {author} {\bibfnamefont {D.}~\bibnamefont {Erben}}, \bibinfo {author} {\bibfnamefont {K.}~\bibnamefont {Tran}}, \bibinfo {author} {\bibfnamefont {D.~S.}\ \bibnamefont {Kim}}, \bibinfo {author} {\bibfnamefont {L.}~\bibnamefont {Sun}}, \bibinfo {author} {\bibfnamefont {J.}~\bibnamefont {Quan}}, \bibinfo {author} {\bibfnamefont {R.}~\bibnamefont {Claassen}}, \bibinfo {author} {\bibfnamefont {S.}~\bibnamefont {Majumder}}, \bibinfo {author} {\bibfnamefont {J.~A.}\ \bibnamefont {Hollingsworth}}, \bibinfo {author} {\bibfnamefont {T.}~\bibnamefont {Taniguchi}}, \bibinfo {author} {\bibfnamefont {K.}~\bibnamefont {Watanabe}}, \bibinfo {author} {\bibfnamefont {K.}~\bibnamefont {Ueno}}, \bibinfo {author} {\bibfnamefont {A.}~\bibnamefont {Singh}}, \bibinfo {author} {\bibfnamefont {G.}~\bibnamefont
  {Moody}}, \bibinfo {author} {\bibfnamefont {F.}~\bibnamefont {Jahnke}},\ and\ \bibinfo {author} {\bibfnamefont {X.}~\bibnamefont {Li}},\ }\bibfield  {title} {\bibinfo {title} {Twist angle-dependent interlayer exciton lifetimes in van der {Waals} heterostructures},\ }\href {https://doi.org/10.1103/PhysRevLett.126.047401} {\bibfield  {journal} {\bibinfo  {journal} {Phys. Rev. Lett.}\ }\textbf {\bibinfo {volume} {126}},\ \bibinfo {pages} {047401} (\bibinfo {year} {2021})}\BibitemShut {NoStop}%
\bibitem [{\citenamefont {Jiang}\ \emph {et~al.}(2018)\citenamefont {Jiang}, \citenamefont {Xu}, \citenamefont {Rasmita}, \citenamefont {Huang}, \citenamefont {Qihua~Xiong},\ and\ \citenamefont {Gao}}]{excitonlifetime_microsecond}%
  \BibitemOpen
  \bibfield  {author} {\bibinfo {author} {\bibfnamefont {C.}~\bibnamefont {Jiang}}, \bibinfo {author} {\bibfnamefont {W.}~\bibnamefont {Xu}}, \bibinfo {author} {\bibfnamefont {A.}~\bibnamefont {Rasmita}}, \bibinfo {author} {\bibfnamefont {Z.}~\bibnamefont {Huang}}, \bibinfo {author} {\bibfnamefont {K.~L.}\ \bibnamefont {Qihua~Xiong}},\ and\ \bibinfo {author} {\bibfnamefont {W.-B.}\ \bibnamefont {Gao}},\ }\bibfield  {title} {\bibinfo {title} {Microsecond dark-exciton valley polarization memory in two-dimensional heterostructures},\ }\href {https://doi.org/10.1038/s41467-018-03174-3} {\bibfield  {journal} {\bibinfo  {journal} {Nature Communications}\ ,\ \bibinfo {pages} {753}} (\bibinfo {year} {2018})}\BibitemShut {NoStop}%
\bibitem [{\citenamefont {Schmitt}\ \emph {et~al.}(2022)\citenamefont {Schmitt}, \citenamefont {Bange}, \citenamefont {Bennecke}, \citenamefont {AlMutairi}, \citenamefont {Meneghini}, \citenamefont {Watanabe}, \citenamefont {Taniguchi}, \citenamefont {Steil}, \citenamefont {Luke}, \citenamefont {Weitz}, \citenamefont {Steil}, \citenamefont {Jansen}, \citenamefont {Brem}, \citenamefont {Malic}, \citenamefont {Hofmann}, \citenamefont {Reutzel},\ and\ \citenamefont {Mathias}}]{giuseppe_nature}%
  \BibitemOpen
  \bibfield  {author} {\bibinfo {author} {\bibfnamefont {D.}~\bibnamefont {Schmitt}}, \bibinfo {author} {\bibfnamefont {J.~P.}\ \bibnamefont {Bange}}, \bibinfo {author} {\bibfnamefont {W.}~\bibnamefont {Bennecke}}, \bibinfo {author} {\bibfnamefont {A.}~\bibnamefont {AlMutairi}}, \bibinfo {author} {\bibfnamefont {G.}~\bibnamefont {Meneghini}}, \bibinfo {author} {\bibfnamefont {K.}~\bibnamefont {Watanabe}}, \bibinfo {author} {\bibfnamefont {T.}~\bibnamefont {Taniguchi}}, \bibinfo {author} {\bibfnamefont {D.}~\bibnamefont {Steil}}, \bibinfo {author} {\bibfnamefont {D.~R.}\ \bibnamefont {Luke}}, \bibinfo {author} {\bibfnamefont {R.~T.}\ \bibnamefont {Weitz}}, \bibinfo {author} {\bibfnamefont {S.}~\bibnamefont {Steil}}, \bibinfo {author} {\bibfnamefont {G.~S.~M.}\ \bibnamefont {Jansen}}, \bibinfo {author} {\bibfnamefont {S.}~\bibnamefont {Brem}}, \bibinfo {author} {\bibfnamefont {E.}~\bibnamefont {Malic}}, \bibinfo {author} {\bibfnamefont {S.}~\bibnamefont {Hofmann}}, \bibinfo {author} {\bibfnamefont
  {M.}~\bibnamefont {Reutzel}},\ and\ \bibinfo {author} {\bibfnamefont {S.}~\bibnamefont {Mathias}},\ }\bibfield  {title} {\bibinfo {title} {Formation of moiré interlayer excitons in space and time},\ }\href {https://doi.org/10.1038/s41586-022-04977-7} {\bibfield  {journal} {\bibinfo  {journal} {Nature}\ ,\ \bibinfo {pages} {499--503}} (\bibinfo {year} {2022})}\BibitemShut {NoStop}%
\bibitem [{\citenamefont {Wang}\ \emph {et~al.}(2018{\natexlab{b}})\citenamefont {Wang}, \citenamefont {Chiu}, \citenamefont {Honz}, \citenamefont {Mak},\ and\ \citenamefont {Shan}}]{mak_interlayer_excitons}%
  \BibitemOpen
  \bibfield  {author} {\bibinfo {author} {\bibfnamefont {Z.}~\bibnamefont {Wang}}, \bibinfo {author} {\bibfnamefont {Y.-H.}\ \bibnamefont {Chiu}}, \bibinfo {author} {\bibfnamefont {K.}~\bibnamefont {Honz}}, \bibinfo {author} {\bibfnamefont {K.~F.}\ \bibnamefont {Mak}},\ and\ \bibinfo {author} {\bibfnamefont {J.}~\bibnamefont {Shan}},\ }\bibfield  {title} {\bibinfo {title} {Electrical tuning of interlayer exciton gases in {WSe}$_2$ bilayers},\ }\href {https://doi.org/10.1021/acs.nanolett.7b03667} {\bibfield  {journal} {\bibinfo  {journal} {Nano Letters}\ }\textbf {\bibinfo {volume} {18}},\ \bibinfo {pages} {137--143} (\bibinfo {year} {2018}{\natexlab{b}})}\BibitemShut {NoStop}%
\bibitem [{\citenamefont {Tagarelli}\ \emph {et~al.}(2023)\citenamefont {Tagarelli}, \citenamefont {Lopriore}, \citenamefont {Erkensten}, \citenamefont {Perea-Causín}, \citenamefont {Brem}, \citenamefont {Hagel}, \citenamefont {Sun}, \citenamefont {Pasquale}, \citenamefont {Watanabe}, \citenamefont {Taniguchi}, \citenamefont {Malic},\ and\ \citenamefont {Kis}}]{tagarelli_natphot}%
  \BibitemOpen
  \bibfield  {author} {\bibinfo {author} {\bibfnamefont {F.}~\bibnamefont {Tagarelli}}, \bibinfo {author} {\bibfnamefont {E.}~\bibnamefont {Lopriore}}, \bibinfo {author} {\bibfnamefont {D.}~\bibnamefont {Erkensten}}, \bibinfo {author} {\bibfnamefont {R.}~\bibnamefont {Perea-Causín}}, \bibinfo {author} {\bibfnamefont {S.}~\bibnamefont {Brem}}, \bibinfo {author} {\bibfnamefont {J.}~\bibnamefont {Hagel}}, \bibinfo {author} {\bibfnamefont {Z.}~\bibnamefont {Sun}}, \bibinfo {author} {\bibfnamefont {G.}~\bibnamefont {Pasquale}}, \bibinfo {author} {\bibfnamefont {K.}~\bibnamefont {Watanabe}}, \bibinfo {author} {\bibfnamefont {T.}~\bibnamefont {Taniguchi}}, \bibinfo {author} {\bibfnamefont {E.}~\bibnamefont {Malic}},\ and\ \bibinfo {author} {\bibfnamefont {A.}~\bibnamefont {Kis}},\ }\bibfield  {title} {\bibinfo {title} {Electrical control of hybrid exciton transport in a van der {Waals} heterostructure},\ }\href {https://doi.org/10.1038/s41566-023-01198-w} {\bibfield  {journal} {\bibinfo  {journal} {Nature
  Photonics}\ ,\ \bibinfo {pages} {615--621}} (\bibinfo {year} {2023})}\BibitemShut {NoStop}%
\bibitem [{\citenamefont {Zhang}\ \emph {et~al.}(2022)\citenamefont {Zhang}, \citenamefont {Regan}, \citenamefont {Wang}, \citenamefont {Zhao}, \citenamefont {Wang}, \citenamefont {Sayyad}, \citenamefont {Yumigeta}, \citenamefont {Watanabe}, \citenamefont {Taniguchi}, \citenamefont {Tongay}, \citenamefont {Crommie}, \citenamefont {Zettl}, \citenamefont {Zaletel},\ and\ \citenamefont {Wang}}]{correlated_exciton_insulator_natphys}%
  \BibitemOpen
  \bibfield  {author} {\bibinfo {author} {\bibfnamefont {Z.}~\bibnamefont {Zhang}}, \bibinfo {author} {\bibfnamefont {E.~C.}\ \bibnamefont {Regan}}, \bibinfo {author} {\bibfnamefont {D.}~\bibnamefont {Wang}}, \bibinfo {author} {\bibfnamefont {W.}~\bibnamefont {Zhao}}, \bibinfo {author} {\bibfnamefont {S.}~\bibnamefont {Wang}}, \bibinfo {author} {\bibfnamefont {M.}~\bibnamefont {Sayyad}}, \bibinfo {author} {\bibfnamefont {K.}~\bibnamefont {Yumigeta}}, \bibinfo {author} {\bibfnamefont {K.}~\bibnamefont {Watanabe}}, \bibinfo {author} {\bibfnamefont {T.}~\bibnamefont {Taniguchi}}, \bibinfo {author} {\bibfnamefont {S.}~\bibnamefont {Tongay}}, \bibinfo {author} {\bibfnamefont {M.}~\bibnamefont {Crommie}}, \bibinfo {author} {\bibfnamefont {A.}~\bibnamefont {Zettl}}, \bibinfo {author} {\bibfnamefont {M.~P.}\ \bibnamefont {Zaletel}},\ and\ \bibinfo {author} {\bibfnamefont {F.}~\bibnamefont {Wang}},\ }\bibfield  {title} {\bibinfo {title} {Correlated interlayer exciton insulator in heterostructures of monolayer
  {WSe}$_2$ and moiré {WS}$_2$/{WSe}$_2$},\ }\href {https://doi.org/10.1038/s41567-022-01702-z} {\bibfield  {journal} {\bibinfo  {journal} {Nature Physics}\ ,\ \bibinfo {pages} {1214--1220}} (\bibinfo {year} {2022})}\BibitemShut {NoStop}%
\bibitem [{\citenamefont {Gu}\ \emph {et~al.}(2022)\citenamefont {Gu}, \citenamefont {Ma}, \citenamefont {Liu}, \citenamefont {Watanabe}, \citenamefont {Taniguchi}, \citenamefont {Hone}, \citenamefont {Shan},\ and\ \citenamefont {Mak}}]{dipolar_exciton_insulator_natphys}%
  \BibitemOpen
  \bibfield  {author} {\bibinfo {author} {\bibfnamefont {J.}~\bibnamefont {Gu}}, \bibinfo {author} {\bibfnamefont {L.}~\bibnamefont {Ma}}, \bibinfo {author} {\bibfnamefont {S.}~\bibnamefont {Liu}}, \bibinfo {author} {\bibfnamefont {K.}~\bibnamefont {Watanabe}}, \bibinfo {author} {\bibfnamefont {T.}~\bibnamefont {Taniguchi}}, \bibinfo {author} {\bibfnamefont {J.~C.}\ \bibnamefont {Hone}}, \bibinfo {author} {\bibfnamefont {J.}~\bibnamefont {Shan}},\ and\ \bibinfo {author} {\bibfnamefont {K.~F.}\ \bibnamefont {Mak}},\ }\bibfield  {title} {\bibinfo {title} {Dipolar excitonic insulator in a moiré lattice},\ }\href {https://doi.org/10.1038/s41567-022-01532-z} {\bibfield  {journal} {\bibinfo  {journal} {Nature Physics}\ ,\ \bibinfo {pages} {395--400}} (\bibinfo {year} {2022})}\BibitemShut {NoStop}%
\bibitem [{\citenamefont {Chen}\ \emph {et~al.}(2022)\citenamefont {Chen}, \citenamefont {Lian}, \citenamefont {Huang}, \citenamefont {Su}, \citenamefont {Rashetnia}, \citenamefont {Ma}, \citenamefont {Yan}, \citenamefont {Blei}, \citenamefont {Xiang}, \citenamefont {Taniguchi}, \citenamefont {Watanabe}, \citenamefont {Tongay}, \citenamefont {Smirnov}, \citenamefont {Wang}, \citenamefont {Zhang}, \citenamefont {Cui},\ and\ \citenamefont {Shi}}]{exciton_insulator_natphys}%
  \BibitemOpen
  \bibfield  {author} {\bibinfo {author} {\bibfnamefont {D.}~\bibnamefont {Chen}}, \bibinfo {author} {\bibfnamefont {Z.}~\bibnamefont {Lian}}, \bibinfo {author} {\bibfnamefont {X.}~\bibnamefont {Huang}}, \bibinfo {author} {\bibfnamefont {Y.}~\bibnamefont {Su}}, \bibinfo {author} {\bibfnamefont {M.}~\bibnamefont {Rashetnia}}, \bibinfo {author} {\bibfnamefont {L.}~\bibnamefont {Ma}}, \bibinfo {author} {\bibfnamefont {L.}~\bibnamefont {Yan}}, \bibinfo {author} {\bibfnamefont {M.}~\bibnamefont {Blei}}, \bibinfo {author} {\bibfnamefont {L.}~\bibnamefont {Xiang}}, \bibinfo {author} {\bibfnamefont {T.}~\bibnamefont {Taniguchi}}, \bibinfo {author} {\bibfnamefont {K.}~\bibnamefont {Watanabe}}, \bibinfo {author} {\bibfnamefont {S.}~\bibnamefont {Tongay}}, \bibinfo {author} {\bibfnamefont {D.}~\bibnamefont {Smirnov}}, \bibinfo {author} {\bibfnamefont {Z.}~\bibnamefont {Wang}}, \bibinfo {author} {\bibfnamefont {C.}~\bibnamefont {Zhang}}, \bibinfo {author} {\bibfnamefont {Y.-T.}\ \bibnamefont {Cui}},\ and\ \bibinfo
  {author} {\bibfnamefont {S.-F.}\ \bibnamefont {Shi}},\ }\bibfield  {title} {\bibinfo {title} {Excitonic insulator in a heterojunction moiré superlattice},\ }\href {https://doi.org/10.1038/s41567-022-01703-y} {\bibfield  {journal} {\bibinfo  {journal} {Nature Physics}\ ,\ \bibinfo {pages} {1171--1176}} (\bibinfo {year} {2022})}\BibitemShut {NoStop}%
\bibitem [{\citenamefont {Xiong}\ \emph {et~al.}(2023)\citenamefont {Xiong}, \citenamefont {Nie}, \citenamefont {Brantly}, \citenamefont {Hays}, \citenamefont {Sailus}, \citenamefont {Watanabe}, \citenamefont {Taniguchi}, \citenamefont {Tongay},\ and\ \citenamefont {Jin}}]{correlated_exciton_insulator_science}%
  \BibitemOpen
  \bibfield  {author} {\bibinfo {author} {\bibfnamefont {R.}~\bibnamefont {Xiong}}, \bibinfo {author} {\bibfnamefont {J.~H.}\ \bibnamefont {Nie}}, \bibinfo {author} {\bibfnamefont {S.~L.}\ \bibnamefont {Brantly}}, \bibinfo {author} {\bibfnamefont {P.}~\bibnamefont {Hays}}, \bibinfo {author} {\bibfnamefont {R.}~\bibnamefont {Sailus}}, \bibinfo {author} {\bibfnamefont {K.}~\bibnamefont {Watanabe}}, \bibinfo {author} {\bibfnamefont {T.}~\bibnamefont {Taniguchi}}, \bibinfo {author} {\bibfnamefont {S.}~\bibnamefont {Tongay}},\ and\ \bibinfo {author} {\bibfnamefont {C.}~\bibnamefont {Jin}},\ }\bibfield  {title} {\bibinfo {title} {Correlated insulator of excitons in {WSe}$_2$/{WS}$_2$ moiré superlattices},\ }\href {https://doi.org/10.1126/science.add5574} {\bibfield  {journal} {\bibinfo  {journal} {Science}\ }\textbf {\bibinfo {volume} {380}},\ \bibinfo {pages} {860--864} (\bibinfo {year} {2023})}\BibitemShut {NoStop}%
\bibitem [{\citenamefont {Wu}\ \emph {et~al.}(2017)\citenamefont {Wu}, \citenamefont {Lovorn},\ and\ \citenamefont {MacDonald}}]{Fengcheng_intralayer_exciton}%
  \BibitemOpen
  \bibfield  {author} {\bibinfo {author} {\bibfnamefont {F.}~\bibnamefont {Wu}}, \bibinfo {author} {\bibfnamefont {T.}~\bibnamefont {Lovorn}},\ and\ \bibinfo {author} {\bibfnamefont {A.~H.}\ \bibnamefont {MacDonald}},\ }\bibfield  {title} {\bibinfo {title} {Topological exciton bands in moir\'e heterojunctions},\ }\href {https://doi.org/10.1103/PhysRevLett.118.147401} {\bibfield  {journal} {\bibinfo  {journal} {Phys. Rev. Lett.}\ }\textbf {\bibinfo {volume} {118}},\ \bibinfo {pages} {147401} (\bibinfo {year} {2017})}\BibitemShut {NoStop}%
\bibitem [{\citenamefont {Xie}\ \emph {et~al.}(2024)\citenamefont {Xie}, \citenamefont {Hafezi},\ and\ \citenamefont {Das~Sarma}}]{dasSarma_topologicalExcitons}%
  \BibitemOpen
  \bibfield  {author} {\bibinfo {author} {\bibfnamefont {M.}~\bibnamefont {Xie}}, \bibinfo {author} {\bibfnamefont {M.}~\bibnamefont {Hafezi}},\ and\ \bibinfo {author} {\bibfnamefont {S.}~\bibnamefont {Das~Sarma}},\ }\bibfield  {title} {\bibinfo {title} {Long-lived topological flatband excitons in semiconductor moir\'e heterostructures: {A} bosonic {Kane-Mele} model platform},\ }\href {https://doi.org/10.1103/PhysRevLett.133.136403} {\bibfield  {journal} {\bibinfo  {journal} {Phys. Rev. Lett.}\ }\textbf {\bibinfo {volume} {133}},\ \bibinfo {pages} {136403} (\bibinfo {year} {2024})}\BibitemShut {NoStop}%
\bibitem [{\citenamefont {Roy}(2014)}]{PhysRevB.90.165139}%
  \BibitemOpen
  \bibfield  {author} {\bibinfo {author} {\bibfnamefont {R.}~\bibnamefont {Roy}},\ }\bibfield  {title} {\bibinfo {title} {Band geometry of fractional topological insulators},\ }\href {https://doi.org/10.1103/PhysRevB.90.165139} {\bibfield  {journal} {\bibinfo  {journal} {Phys. Rev. B}\ }\textbf {\bibinfo {volume} {90}},\ \bibinfo {pages} {165139} (\bibinfo {year} {2014})}\BibitemShut {NoStop}%
\bibitem [{\citenamefont {Wang}\ \emph {et~al.}(2021)\citenamefont {Wang}, \citenamefont {Cano}, \citenamefont {Millis}, \citenamefont {Liu},\ and\ \citenamefont {Yang}}]{jie_wang_qgt2}%
  \BibitemOpen
  \bibfield  {author} {\bibinfo {author} {\bibfnamefont {J.}~\bibnamefont {Wang}}, \bibinfo {author} {\bibfnamefont {J.}~\bibnamefont {Cano}}, \bibinfo {author} {\bibfnamefont {A.~J.}\ \bibnamefont {Millis}}, \bibinfo {author} {\bibfnamefont {Z.}~\bibnamefont {Liu}},\ and\ \bibinfo {author} {\bibfnamefont {B.}~\bibnamefont {Yang}},\ }\bibfield  {title} {\bibinfo {title} {Exact {Landau} level description of geometry and interaction in a flatband},\ }\href {https://doi.org/10.1103/PhysRevLett.127.246403} {\bibfield  {journal} {\bibinfo  {journal} {Phys. Rev. Lett.}\ }\textbf {\bibinfo {volume} {127}},\ \bibinfo {pages} {246403} (\bibinfo {year} {2021})}\BibitemShut {NoStop}%
\bibitem [{\citenamefont {Ledwith}\ \emph {et~al.}(2023)\citenamefont {Ledwith}, \citenamefont {Vishwanath},\ and\ \citenamefont {Parker}}]{Vortexability_band}%
  \BibitemOpen
  \bibfield  {author} {\bibinfo {author} {\bibfnamefont {P.~J.}\ \bibnamefont {Ledwith}}, \bibinfo {author} {\bibfnamefont {A.}~\bibnamefont {Vishwanath}},\ and\ \bibinfo {author} {\bibfnamefont {D.~E.}\ \bibnamefont {Parker}},\ }\bibfield  {title} {\bibinfo {title} {Vortexability: A unifying criterion for ideal fractional {Chern} insulators},\ }\href {https://doi.org/10.1103/PhysRevB.108.205144} {\bibfield  {journal} {\bibinfo  {journal} {Phys. Rev. B}\ }\textbf {\bibinfo {volume} {108}},\ \bibinfo {pages} {205144} (\bibinfo {year} {2023})}\BibitemShut {NoStop}%
\bibitem [{\citenamefont {Liu}\ \emph {et~al.}(2025{\natexlab{c}})\citenamefont {Liu}, \citenamefont {Yang}, \citenamefont {Abouelkomsan}, \citenamefont {Liu},\ and\ \citenamefont {Bergholtz}}]{Hui_broken_symmetry}%
  \BibitemOpen
  \bibfield  {author} {\bibinfo {author} {\bibfnamefont {H.}~\bibnamefont {Liu}}, \bibinfo {author} {\bibfnamefont {K.}~\bibnamefont {Yang}}, \bibinfo {author} {\bibfnamefont {A.}~\bibnamefont {Abouelkomsan}}, \bibinfo {author} {\bibfnamefont {Z.}~\bibnamefont {Liu}},\ and\ \bibinfo {author} {\bibfnamefont {E.~J.}\ \bibnamefont {Bergholtz}},\ }\bibfield  {title} {\bibinfo {title} {Broken symmetry in ideal {Chern} bands},\ }\href {https://link.aps.org/doi/10.1103/PhysRevB.111.L201105} {\bibfield  {journal} {\bibinfo  {journal} {Phys. Rev. B}\ }\textbf {\bibinfo {volume} {111}},\ \bibinfo {pages} {L201105} (\bibinfo {year} {2025}{\natexlab{c}})}\BibitemShut {NoStop}%
\bibitem [{\citenamefont {Perea-Causin}\ \emph {et~al.}(2025)\citenamefont {Perea-Causin}, \citenamefont {Liu},\ and\ \citenamefont {Bergholtz}}]{raulQAHC}%
  \BibitemOpen
  \bibfield  {author} {\bibinfo {author} {\bibfnamefont {R.}~\bibnamefont {Perea-Causin}}, \bibinfo {author} {\bibfnamefont {H.}~\bibnamefont {Liu}},\ and\ \bibinfo {author} {\bibfnamefont {E.~J.}\ \bibnamefont {Bergholtz}},\ }\bibfield  {title} {\bibinfo {title} {Quantum anomalous {Hall} crystals in moir\'e bands with higher {Chern} number},\ }\href {https://doi.org/10.1038/s41467-025-62224-9} {\bibfield  {journal} {\bibinfo  {journal} {Nature Communications}\ }\textbf {\bibinfo {volume} {16}},\ \bibinfo {pages} {6875} (\bibinfo {year} {2025})}\BibitemShut {NoStop}%
\bibitem [{\citenamefont {Ji}\ and\ \citenamefont {Yang}(2024)}]{ji2024quantummetricinducedhole}%
  \BibitemOpen
  \bibfield  {author} {\bibinfo {author} {\bibfnamefont {G.}~\bibnamefont {Ji}}\ and\ \bibinfo {author} {\bibfnamefont {B.}~\bibnamefont {Yang}},\ }\href {https://arxiv.org/abs/2409.08324} {\bibinfo {title} {Quantum metric induced hole dispersion and emergent particle-hole symmetry in topological flat bands}} (\bibinfo {year} {2024}),\ \Eprint {https://arxiv.org/abs/2409.08324} {arXiv:2409.08324 [cond-mat.str-el]} \BibitemShut {NoStop}%
\bibitem [{\citenamefont {Simon}\ \emph {et~al.}(2015)\citenamefont {Simon}, \citenamefont {Harper},\ and\ \citenamefont {Read}}]{simonFCIzeroBerry}%
  \BibitemOpen
  \bibfield  {author} {\bibinfo {author} {\bibfnamefont {S.~H.}\ \bibnamefont {Simon}}, \bibinfo {author} {\bibfnamefont {F.}~\bibnamefont {Harper}},\ and\ \bibinfo {author} {\bibfnamefont {N.}~\bibnamefont {Read}},\ }\bibfield  {title} {\bibinfo {title} {Fractional {Chern} insulators in bands with zero {Berry} curvature},\ }\href {https://doi.org/10.1103/PhysRevB.92.195104} {\bibfield  {journal} {\bibinfo  {journal} {Phys. Rev. B}\ }\textbf {\bibinfo {volume} {92}},\ \bibinfo {pages} {195104} (\bibinfo {year} {2015})}\BibitemShut {NoStop}%
\bibitem [{\citenamefont {Ma}\ \emph {et~al.}(2021)\citenamefont {Ma}, \citenamefont {Nguyen}, \citenamefont {Wang}, \citenamefont {Zeng}, \citenamefont {Watanabe}, \citenamefont {Taniguchi}, \citenamefont {MacDonald}, \citenamefont {Mak},\ and\ \citenamefont {Shan}}]{exciton_insulator_nat}%
  \BibitemOpen
  \bibfield  {author} {\bibinfo {author} {\bibfnamefont {L.}~\bibnamefont {Ma}}, \bibinfo {author} {\bibfnamefont {P.~X.}\ \bibnamefont {Nguyen}}, \bibinfo {author} {\bibfnamefont {Z.}~\bibnamefont {Wang}}, \bibinfo {author} {\bibfnamefont {Y.}~\bibnamefont {Zeng}}, \bibinfo {author} {\bibfnamefont {K.}~\bibnamefont {Watanabe}}, \bibinfo {author} {\bibfnamefont {T.}~\bibnamefont {Taniguchi}}, \bibinfo {author} {\bibfnamefont {A.~H.}\ \bibnamefont {MacDonald}}, \bibinfo {author} {\bibfnamefont {K.~F.}\ \bibnamefont {Mak}},\ and\ \bibinfo {author} {\bibfnamefont {J.}~\bibnamefont {Shan}},\ }\bibfield  {title} {\bibinfo {title} {Strongly correlated excitonic insulator in atomic double layers},\ }\href {https://doi.org/10.1038/s41586-021-03947-9} {\bibfield  {journal} {\bibinfo  {journal} {Nature}\ ,\ \bibinfo {pages} {585--589}} (\bibinfo {year} {2021})}\BibitemShut {NoStop}%
\bibitem [{Sup()}]{SupMat}%
  \BibitemOpen
  \href@noop {} {\bibinfo  {journal} {See the Supplemental Material for additional details.}\ }\BibitemShut {NoStop}%
\bibitem [{\citenamefont {Fukui}\ \emph {et~al.}(2005)\citenamefont {Fukui}, \citenamefont {Hatsugai},\ and\ \citenamefont {Suzuki}}]{Fukui_Chern}%
  \BibitemOpen
\bibfield  {journal} {  }\bibfield  {author} {\bibinfo {author} {\bibfnamefont {T.}~\bibnamefont {Fukui}}, \bibinfo {author} {\bibfnamefont {Y.}~\bibnamefont {Hatsugai}},\ and\ \bibinfo {author} {\bibfnamefont {H.}~\bibnamefont {Suzuki}},\ }\bibfield  {title} {\bibinfo {title} {Chern numbers in discretized {Brillouin} zone: Efficient method of computing (spin) {Hall} conductances},\ }\href {https://doi.org/10.1143/JPSJ.74.1674} {\bibfield  {journal} {\bibinfo  {journal} {Journal of the Physical Society of Japan}\ }\textbf {\bibinfo {volume} {74}},\ \bibinfo {pages} {1674--1677} (\bibinfo {year} {2005})}\BibitemShut {NoStop}%
\bibitem [{\citenamefont {Repellin}\ \emph {et~al.}(2014)\citenamefont {Repellin}, \citenamefont {Bernevig},\ and\ \citenamefont {Regnault}}]{Repellin_tilted_sample}%
  \BibitemOpen
  \bibfield  {author} {\bibinfo {author} {\bibfnamefont {C.}~\bibnamefont {Repellin}}, \bibinfo {author} {\bibfnamefont {B.~A.}\ \bibnamefont {Bernevig}},\ and\ \bibinfo {author} {\bibfnamefont {N.}~\bibnamefont {Regnault}},\ }\bibfield  {title} {\bibinfo {title} {$\mathbb{Z}_{2}$ fractional topological insulators in two dimensions},\ }\href {https://doi.org/10.1103/PhysRevB.90.245401} {\bibfield  {journal} {\bibinfo  {journal} {Phys. Rev. B}\ }\textbf {\bibinfo {volume} {90}},\ \bibinfo {pages} {245401} (\bibinfo {year} {2014})}\BibitemShut {NoStop}%
\bibitem [{\citenamefont {Brem}\ \emph {et~al.}(2020)\citenamefont {Brem}, \citenamefont {Linderälv}, \citenamefont {Erhart},\ and\ \citenamefont {Malic}}]{brem_moire}%
  \BibitemOpen
  \bibfield  {author} {\bibinfo {author} {\bibfnamefont {S.}~\bibnamefont {Brem}}, \bibinfo {author} {\bibfnamefont {C.}~\bibnamefont {Linderälv}}, \bibinfo {author} {\bibfnamefont {P.}~\bibnamefont {Erhart}},\ and\ \bibinfo {author} {\bibfnamefont {E.}~\bibnamefont {Malic}},\ }\bibfield  {title} {\bibinfo {title} {Tunable phases of moiré excitons in van der {Waals} heterostructures},\ }\href {https://doi.org/10.1021/acs.nanolett.0c03019} {\bibfield  {journal} {\bibinfo  {journal} {Nano Letters}\ }\textbf {\bibinfo {volume} {20}},\ \bibinfo {pages} {8534--8540} (\bibinfo {year} {2020})}\BibitemShut {NoStop}%
\bibitem [{\citenamefont {Jankowski}\ \emph {et~al.}(2025)\citenamefont {Jankowski}, \citenamefont {Thompson}, \citenamefont {Monserrat},\ and\ \citenamefont {Slager}}]{wojciech2025quantummetric}%
  \BibitemOpen
  \bibfield  {author} {\bibinfo {author} {\bibfnamefont {W.~J.}\ \bibnamefont {Jankowski}}, \bibinfo {author} {\bibfnamefont {J.~J.~P.}\ \bibnamefont {Thompson}}, \bibinfo {author} {\bibfnamefont {B.}~\bibnamefont {Monserrat}},\ and\ \bibinfo {author} {\bibfnamefont {R.-J.}\ \bibnamefont {Slager}},\ }\bibfield  {title} {\bibinfo {title} {Excitonic topology and quantum geometry in organic semiconductors},\ }\href {https://doi.org/10.1038/s41467-025-59257-5} {\bibfield  {journal} {\bibinfo  {journal} {Nature Communications}\ }\textbf {\bibinfo {volume} {16}},\ \bibinfo {pages} {4661} (\bibinfo {year} {2025})}\BibitemShut {NoStop}%
\bibitem [{\citenamefont {Tarnopolsky}\ \emph {et~al.}(2019)\citenamefont {Tarnopolsky}, \citenamefont {Kruchkov},\ and\ \citenamefont {Vishwanath}}]{ideal_band_tbg}%
  \BibitemOpen
  \bibfield  {author} {\bibinfo {author} {\bibfnamefont {G.}~\bibnamefont {Tarnopolsky}}, \bibinfo {author} {\bibfnamefont {A.~J.}\ \bibnamefont {Kruchkov}},\ and\ \bibinfo {author} {\bibfnamefont {A.}~\bibnamefont {Vishwanath}},\ }\bibfield  {title} {\bibinfo {title} {Origin of magic angles in twisted bilayer graphene},\ }\href {https://doi.org/10.1103/PhysRevLett.122.106405} {\bibfield  {journal} {\bibinfo  {journal} {Phys. Rev. Lett.}\ }\textbf {\bibinfo {volume} {122}},\ \bibinfo {pages} {106405} (\bibinfo {year} {2019})}\BibitemShut {NoStop}%
\bibitem [{\citenamefont {Abouelkomsan}\ \emph {et~al.}(2023)\citenamefont {Abouelkomsan}, \citenamefont {Yang},\ and\ \citenamefont {Bergholtz}}]{PhysRevResearch.5.L012015}%
  \BibitemOpen
  \bibfield  {author} {\bibinfo {author} {\bibfnamefont {A.}~\bibnamefont {Abouelkomsan}}, \bibinfo {author} {\bibfnamefont {K.}~\bibnamefont {Yang}},\ and\ \bibinfo {author} {\bibfnamefont {E.~J.}\ \bibnamefont {Bergholtz}},\ }\bibfield  {title} {\bibinfo {title} {Quantum metric induced phases in moir\'e materials},\ }\href {https://doi.org/10.1103/PhysRevResearch.5.L012015} {\bibfield  {journal} {\bibinfo  {journal} {Phys. Rev. Res.}\ }\textbf {\bibinfo {volume} {5}},\ \bibinfo {pages} {L012015} (\bibinfo {year} {2023})}\BibitemShut {NoStop}%
\bibitem [{\citenamefont {Haug}\ and\ \citenamefont {Schmitt-Rink}(1984)}]{HAUG19843}%
  \BibitemOpen
  \bibfield  {author} {\bibinfo {author} {\bibfnamefont {H.}~\bibnamefont {Haug}}\ and\ \bibinfo {author} {\bibfnamefont {S.}~\bibnamefont {Schmitt-Rink}},\ }\bibfield  {title} {\bibinfo {title} {Electron theory of the optical properties of laser-excited semiconductors},\ }\href {https://doi.org/https://doi.org/10.1016/0079-6727(84)90026-0} {\bibfield  {journal} {\bibinfo  {journal} {Progress in Quantum Electronics}\ }\textbf {\bibinfo {volume} {9}},\ \bibinfo {pages} {3--100} (\bibinfo {year} {1984})}\BibitemShut {NoStop}%
\bibitem [{\citenamefont {Laughlin}(1983)}]{Laughlin}%
  \BibitemOpen
  \bibfield  {author} {\bibinfo {author} {\bibfnamefont {R.~B.}\ \bibnamefont {Laughlin}},\ }\bibfield  {title} {\bibinfo {title} {Anomalous quantum {Hall} effect: An incompressible quantum fluid with fractionally charged excitations},\ }\href {https://doi.org/10.1103/PhysRevLett.50.1395} {\bibfield  {journal} {\bibinfo  {journal} {Phys. Rev. Lett.}\ }\textbf {\bibinfo {volume} {50}},\ \bibinfo {pages} {1395--1398} (\bibinfo {year} {1983})}\BibitemShut {NoStop}%
\bibitem [{\citenamefont {Haldane}(1983)}]{haldane_fqhe}%
  \BibitemOpen
  \bibfield  {author} {\bibinfo {author} {\bibfnamefont {F.~D.~M.}\ \bibnamefont {Haldane}},\ }\bibfield  {title} {\bibinfo {title} {Fractional quantization of the {Hall} effect: A hierarchy of incompressible quantum fluid states},\ }\href {https://doi.org/10.1103/PhysRevLett.51.605} {\bibfield  {journal} {\bibinfo  {journal} {Phys. Rev. Lett.}\ }\textbf {\bibinfo {volume} {51}},\ \bibinfo {pages} {605--608} (\bibinfo {year} {1983})}\BibitemShut {NoStop}%
\bibitem [{\citenamefont {Kwan}\ \emph {et~al.}(2022)\citenamefont {Kwan}, \citenamefont {Hu}, \citenamefont {Simon},\ and\ \citenamefont {Parameswaran}}]{oxford_fci_exciton}%
  \BibitemOpen
  \bibfield  {author} {\bibinfo {author} {\bibfnamefont {Y.~H.}\ \bibnamefont {Kwan}}, \bibinfo {author} {\bibfnamefont {Y.}~\bibnamefont {Hu}}, \bibinfo {author} {\bibfnamefont {S.~H.}\ \bibnamefont {Simon}},\ and\ \bibinfo {author} {\bibfnamefont {S.~A.}\ \bibnamefont {Parameswaran}},\ }\bibfield  {title} {\bibinfo {title} {Excitonic fractional quantum {Hall} hierarchy in moir\'e heterostructures},\ }\href {https://doi.org/10.1103/PhysRevB.105.235121} {\bibfield  {journal} {\bibinfo  {journal} {Phys. Rev. B}\ }\textbf {\bibinfo {volume} {105}},\ \bibinfo {pages} {235121} (\bibinfo {year} {2022})}\BibitemShut {NoStop}%
\bibitem [{\citenamefont {Stefanidis}\ and\ \citenamefont {Sodemann}(2020)}]{sodemann_fci_exciton}%
  \BibitemOpen
  \bibfield  {author} {\bibinfo {author} {\bibfnamefont {N.}~\bibnamefont {Stefanidis}}\ and\ \bibinfo {author} {\bibfnamefont {I.}~\bibnamefont {Sodemann}},\ }\bibfield  {title} {\bibinfo {title} {Excitonic {Laughlin} states in ideal topological insulator flat bands and their possible presence in moir\'e superlattice materials},\ }\href {https://doi.org/10.1103/PhysRevB.102.035158} {\bibfield  {journal} {\bibinfo  {journal} {Phys. Rev. B}\ }\textbf {\bibinfo {volume} {102}},\ \bibinfo {pages} {035158} (\bibinfo {year} {2020})}\BibitemShut {NoStop}%
\bibitem [{\citenamefont {Hu}\ \emph {et~al.}(2018)\citenamefont {Hu}, \citenamefont {Venderbos},\ and\ \citenamefont {Kane}}]{Kane_fci_exciton}%
  \BibitemOpen
  \bibfield  {author} {\bibinfo {author} {\bibfnamefont {Y.}~\bibnamefont {Hu}}, \bibinfo {author} {\bibfnamefont {J.~W.~F.}\ \bibnamefont {Venderbos}},\ and\ \bibinfo {author} {\bibfnamefont {C.~L.}\ \bibnamefont {Kane}},\ }\bibfield  {title} {\bibinfo {title} {Fractional excitonic insulator},\ }\href {https://doi.org/10.1103/PhysRevLett.121.126601} {\bibfield  {journal} {\bibinfo  {journal} {Phys. Rev. Lett.}\ }\textbf {\bibinfo {volume} {121}},\ \bibinfo {pages} {126601} (\bibinfo {year} {2018})}\BibitemShut {NoStop}%
\bibitem [{\citenamefont {Bergholtz}\ and\ \citenamefont {Karlhede}(2005)}]{bergholtz2005half}%
  \BibitemOpen
  \bibfield  {author} {\bibinfo {author} {\bibfnamefont {E.~J.}\ \bibnamefont {Bergholtz}}\ and\ \bibinfo {author} {\bibfnamefont {A.}~\bibnamefont {Karlhede}},\ }\bibfield  {title} {\bibinfo {title} {Half-filled lowest {Landau} level on a thin torus},\ }\href {https://doi.org/10.1103/PhysRevLett.94.026802} {\bibfield  {journal} {\bibinfo  {journal} {Phys. Rev. Lett.}\ }\textbf {\bibinfo {volume} {94}},\ \bibinfo {pages} {026802} (\bibinfo {year} {2005})}\BibitemShut {NoStop}%
\bibitem [{\citenamefont {Ardonne}\ \emph {et~al.}(2008)\citenamefont {Ardonne}, \citenamefont {Bergholtz}, \citenamefont {Kailasvuori},\ and\ \citenamefont {Wikberg}}]{ardonne2008degeneracy}%
  \BibitemOpen
  \bibfield  {author} {\bibinfo {author} {\bibfnamefont {E.}~\bibnamefont {Ardonne}}, \bibinfo {author} {\bibfnamefont {E.~J.}\ \bibnamefont {Bergholtz}}, \bibinfo {author} {\bibfnamefont {J.}~\bibnamefont {Kailasvuori}},\ and\ \bibinfo {author} {\bibfnamefont {E.}~\bibnamefont {Wikberg}},\ }\bibfield  {title} {\bibinfo {title} {Degeneracy of non-{Abelian} quantum {Hall} states on the torus: domain walls and conformal field theory},\ }\href {https://doi.org/10.1088/1742-5468/2008/04/P04016} {\bibfield  {journal} {\bibinfo  {journal} {Journal of Statistical Mechanics: Theory and Experiment}\ }\textbf {\bibinfo {volume} {2008}},\ \bibinfo {pages} {P04016} (\bibinfo {year} {2008})}\BibitemShut {NoStop}%
\bibitem [{\citenamefont {Erkensten}\ \emph {et~al.}(2022)\citenamefont {Erkensten}, \citenamefont {Brem}, \citenamefont {Perea-Caus\'{\i}n},\ and\ \citenamefont {Malic}}]{erkensten_interlayerExcitons}%
  \BibitemOpen
  \bibfield  {author} {\bibinfo {author} {\bibfnamefont {D.}~\bibnamefont {Erkensten}}, \bibinfo {author} {\bibfnamefont {S.}~\bibnamefont {Brem}}, \bibinfo {author} {\bibfnamefont {R.}~\bibnamefont {Perea-Caus\'{\i}n}},\ and\ \bibinfo {author} {\bibfnamefont {E.}~\bibnamefont {Malic}},\ }\bibfield  {title} {\bibinfo {title} {Microscopic origin of anomalous interlayer exciton transport in van der {Waals} heterostructures},\ }\href {https://doi.org/10.1103/PhysRevMaterials.6.094006} {\bibfield  {journal} {\bibinfo  {journal} {Phys. Rev. Mater.}\ }\textbf {\bibinfo {volume} {6}},\ \bibinfo {pages} {094006} (\bibinfo {year} {2022})}\BibitemShut {NoStop}%
\bibitem [{\citenamefont {Steinhoff}\ \emph {et~al.}(2024)\citenamefont {Steinhoff}, \citenamefont {Wietek}, \citenamefont {Florian}, \citenamefont {Schulz}, \citenamefont {Taniguchi}, \citenamefont {Watanabe}, \citenamefont {Zhao}, \citenamefont {H\"ogele}, \citenamefont {Jahnke},\ and\ \citenamefont {Chernikov}}]{steinhoff_xxinteractions}%
  \BibitemOpen
  \bibfield  {author} {\bibinfo {author} {\bibfnamefont {A.}~\bibnamefont {Steinhoff}}, \bibinfo {author} {\bibfnamefont {E.}~\bibnamefont {Wietek}}, \bibinfo {author} {\bibfnamefont {M.}~\bibnamefont {Florian}}, \bibinfo {author} {\bibfnamefont {T.}~\bibnamefont {Schulz}}, \bibinfo {author} {\bibfnamefont {T.}~\bibnamefont {Taniguchi}}, \bibinfo {author} {\bibfnamefont {K.}~\bibnamefont {Watanabe}}, \bibinfo {author} {\bibfnamefont {S.}~\bibnamefont {Zhao}}, \bibinfo {author} {\bibfnamefont {A.}~\bibnamefont {H\"ogele}}, \bibinfo {author} {\bibfnamefont {F.}~\bibnamefont {Jahnke}},\ and\ \bibinfo {author} {\bibfnamefont {A.}~\bibnamefont {Chernikov}},\ }\bibfield  {title} {\bibinfo {title} {Exciton-exciton interactions in van der {Waals} heterobilayers},\ }\href {https://doi.org/10.1103/PhysRevX.14.031025} {\bibfield  {journal} {\bibinfo  {journal} {Phys. Rev. X}\ }\textbf {\bibinfo {volume} {14}},\ \bibinfo {pages} {031025} (\bibinfo {year} {2024})}\BibitemShut {NoStop}%
\bibitem [{\citenamefont {Kyriienko}\ \emph {et~al.}(2012)\citenamefont {Kyriienko}, \citenamefont {Magnusson},\ and\ \citenamefont {Shelykh}}]{kyriienko_excitons}%
  \BibitemOpen
  \bibfield  {author} {\bibinfo {author} {\bibfnamefont {O.}~\bibnamefont {Kyriienko}}, \bibinfo {author} {\bibfnamefont {E.~B.}\ \bibnamefont {Magnusson}},\ and\ \bibinfo {author} {\bibfnamefont {I.~A.}\ \bibnamefont {Shelykh}},\ }\bibfield  {title} {\bibinfo {title} {Spin dynamics of cold exciton condensates},\ }\href {https://doi.org/10.1103/PhysRevB.86.115324} {\bibfield  {journal} {\bibinfo  {journal} {Phys. Rev. B}\ }\textbf {\bibinfo {volume} {86}},\ \bibinfo {pages} {115324} (\bibinfo {year} {2012})}\BibitemShut {NoStop}%
\bibitem [{\citenamefont {Brem}\ and\ \citenamefont {Malic}(2023)}]{samuel_bosonic}%
  \BibitemOpen
  \bibfield  {author} {\bibinfo {author} {\bibfnamefont {S.}~\bibnamefont {Brem}}\ and\ \bibinfo {author} {\bibfnamefont {E.}~\bibnamefont {Malic}},\ }\bibfield  {title} {\bibinfo {title} {Bosonic delocalization of dipolar moiré excitons},\ }\href {https://doi.org/10.1021/acs.nanolett.3c01160} {\bibfield  {journal} {\bibinfo  {journal} {Nano Letters}\ }\textbf {\bibinfo {volume} {23}},\ \bibinfo {pages} {4627--4633} (\bibinfo {year} {2023})}\BibitemShut {NoStop}%
\bibitem [{\citenamefont {Reddy}\ \emph {et~al.}(2023)\citenamefont {Reddy}, \citenamefont {Alsallom}, \citenamefont {Zhang}, \citenamefont {Devakul},\ and\ \citenamefont {Fu}}]{PhysRevB.108.085117}%
  \BibitemOpen
  \bibfield  {author} {\bibinfo {author} {\bibfnamefont {A.~P.}\ \bibnamefont {Reddy}}, \bibinfo {author} {\bibfnamefont {F.}~\bibnamefont {Alsallom}}, \bibinfo {author} {\bibfnamefont {Y.}~\bibnamefont {Zhang}}, \bibinfo {author} {\bibfnamefont {T.}~\bibnamefont {Devakul}},\ and\ \bibinfo {author} {\bibfnamefont {L.}~\bibnamefont {Fu}},\ }\bibfield  {title} {\bibinfo {title} {Fractional quantum anomalous {Hall} states in twisted bilayer ${\mathrm{mote}}_{2}$ and ${\mathrm{wse}}_{2}$},\ }\href {https://doi.org/10.1103/PhysRevB.108.085117} {\bibfield  {journal} {\bibinfo  {journal} {Phys. Rev. B}\ }\textbf {\bibinfo {volume} {108}},\ \bibinfo {pages} {085117} (\bibinfo {year} {2023})}\BibitemShut {NoStop}%
\bibitem [{\citenamefont {{Park}}\ \emph {et~al.}(2025)\citenamefont {{Park}}, \citenamefont {{Li}}, \citenamefont {{Hu}}, \citenamefont {{Beach}}, \citenamefont {{Gon{\c{c}}alves}}, \citenamefont {{Mendez-Valderrama}}, \citenamefont {{Herzog-Arbeitman}}, \citenamefont {{Taniguchi}}, \citenamefont {{Watanabe}}, \citenamefont {{Cobden}}, \citenamefont {{Fu}}, \citenamefont {{Bernevig}}, \citenamefont {{Regnault}}, \citenamefont {{Chu}}, \citenamefont {{Xiao}},\ and\ \citenamefont {{Xu}}}]{xiaodong_high_temp_fci}%
  \BibitemOpen
  \bibfield  {author} {\bibinfo {author} {\bibfnamefont {H.}~\bibnamefont {{Park}}}, \bibinfo {author} {\bibfnamefont {W.}~\bibnamefont {{Li}}}, \bibinfo {author} {\bibfnamefont {C.}~\bibnamefont {{Hu}}}, \bibinfo {author} {\bibfnamefont {C.}~\bibnamefont {{Beach}}}, \bibinfo {author} {\bibfnamefont {M.}~\bibnamefont {{Gon{\c{c}}alves}}}, \bibinfo {author} {\bibfnamefont {J.~F.}\ \bibnamefont {{Mendez-Valderrama}}}, \bibinfo {author} {\bibfnamefont {J.}~\bibnamefont {{Herzog-Arbeitman}}}, \bibinfo {author} {\bibfnamefont {T.}~\bibnamefont {{Taniguchi}}}, \bibinfo {author} {\bibfnamefont {K.}~\bibnamefont {{Watanabe}}}, \bibinfo {author} {\bibfnamefont {D.}~\bibnamefont {{Cobden}}}, \bibinfo {author} {\bibfnamefont {L.}~\bibnamefont {{Fu}}}, \bibinfo {author} {\bibfnamefont {B.~A.}\ \bibnamefont {{Bernevig}}}, \bibinfo {author} {\bibfnamefont {N.}~\bibnamefont {{Regnault}}}, \bibinfo {author} {\bibfnamefont {J.-H.}\ \bibnamefont {{Chu}}}, \bibinfo {author} {\bibfnamefont {D.}~\bibnamefont {{Xiao}}},\ and\
  \bibinfo {author} {\bibfnamefont {X.}~\bibnamefont {{Xu}}},\ }\bibfield  {title} {\bibinfo {title} {{Observation of High-Temperature Dissipationless Fractional {Chern} Insulator}},\ }\href {https://doi.org/10.48550/arXiv.2503.10989} {\bibfield  {journal} {\bibinfo  {journal} {arXiv e-prints}\ ,\ \bibinfo {eid} {arXiv:2503.10989}} (\bibinfo {year} {2025})},\ \Eprint {https://arxiv.org/abs/2503.10989} {arXiv:2503.10989 [cond-mat.mes-hall]} \BibitemShut {NoStop}%
\bibitem [{\citenamefont {Moore}\ and\ \citenamefont {Read}(1991)}]{MOORE1991362}%
  \BibitemOpen
  \bibfield  {author} {\bibinfo {author} {\bibfnamefont {G.}~\bibnamefont {Moore}}\ and\ \bibinfo {author} {\bibfnamefont {N.}~\bibnamefont {Read}},\ }\bibfield  {title} {\bibinfo {title} {Nonabelions in the fractional quantum {Hall} effect},\ }\href {https://doi.org/https://doi.org/10.1016/0550-3213(91)90407-O} {\bibfield  {journal} {\bibinfo  {journal} {Nuclear Physics B}\ }\textbf {\bibinfo {volume} {360}},\ \bibinfo {pages} {362--396} (\bibinfo {year} {1991})}\BibitemShut {NoStop}%
\bibitem [{\citenamefont {Greiter}\ \emph {et~al.}(1992)\citenamefont {Greiter}, \citenamefont {Wen},\ and\ \citenamefont {Wilczek}}]{greiter1992}%
  \BibitemOpen
  \bibfield  {author} {\bibinfo {author} {\bibfnamefont {M.}~\bibnamefont {Greiter}}, \bibinfo {author} {\bibfnamefont {X.-G.}\ \bibnamefont {Wen}},\ and\ \bibinfo {author} {\bibfnamefont {F.}~\bibnamefont {Wilczek}},\ }\bibfield  {title} {\bibinfo {title} {Paired {Hall} states},\ }\href {https://www.sciencedirect.com/science/article/pii/055032139290401V} {\bibfield  {journal} {\bibinfo  {journal} {Nuclear Physics B}\ }\textbf {\bibinfo {volume} {374}},\ \bibinfo {pages} {567--614} (\bibinfo {year} {1992})}\BibitemShut {NoStop}%
\bibitem [{\citenamefont {Cooper}\ \emph {et~al.}(2001)\citenamefont {Cooper}, \citenamefont {Wilkin},\ and\ \citenamefont {Gunn}}]{cooper_rotatingBEC}%
  \BibitemOpen
  \bibfield  {author} {\bibinfo {author} {\bibfnamefont {N.~R.}\ \bibnamefont {Cooper}}, \bibinfo {author} {\bibfnamefont {N.~K.}\ \bibnamefont {Wilkin}},\ and\ \bibinfo {author} {\bibfnamefont {J.~M.~F.}\ \bibnamefont {Gunn}},\ }\bibfield  {title} {\bibinfo {title} {Quantum phases of vortices in rotating {Bose-Einstein} condensates},\ }\href {https://doi.org/10.1103/PhysRevLett.87.120405} {\bibfield  {journal} {\bibinfo  {journal} {Phys. Rev. Lett.}\ }\textbf {\bibinfo {volume} {87}},\ \bibinfo {pages} {120405} (\bibinfo {year} {2001})}\BibitemShut {NoStop}%
\bibitem [{\citenamefont {Regnault}\ and\ \citenamefont {Jolicoeur}(2003)}]{regnault_rotatingBEC}%
  \BibitemOpen
  \bibfield  {author} {\bibinfo {author} {\bibfnamefont {N.}~\bibnamefont {Regnault}}\ and\ \bibinfo {author} {\bibfnamefont {T.}~\bibnamefont {Jolicoeur}},\ }\bibfield  {title} {\bibinfo {title} {Quantum {Hall} fractions in rotating {Bose-Einstein} condensates},\ }\href {https://doi.org/10.1103/PhysRevLett.91.030402} {\bibfield  {journal} {\bibinfo  {journal} {Phys. Rev. Lett.}\ }\textbf {\bibinfo {volume} {91}},\ \bibinfo {pages} {030402} (\bibinfo {year} {2003})}\BibitemShut {NoStop}%
\bibitem [{\citenamefont {Liu}\ \emph {et~al.}(2013)\citenamefont {Liu}, \citenamefont {Bergholtz},\ and\ \citenamefont {Kapit}}]{NAlong2013}%
  \BibitemOpen
  \bibfield  {author} {\bibinfo {author} {\bibfnamefont {Z.}~\bibnamefont {Liu}}, \bibinfo {author} {\bibfnamefont {E.~J.}\ \bibnamefont {Bergholtz}},\ and\ \bibinfo {author} {\bibfnamefont {E.}~\bibnamefont {Kapit}},\ }\bibfield  {title} {\bibinfo {title} {Non-{Abelian} fractional {Chern} insulators from long-range interactions},\ }\href {https://doi.org/10.1103/PhysRevB.88.205101} {\bibfield  {journal} {\bibinfo  {journal} {Phys. Rev. B}\ }\textbf {\bibinfo {volume} {88}},\ \bibinfo {pages} {205101} (\bibinfo {year} {2013})}\BibitemShut {NoStop}%
\bibitem [{\citenamefont {Bergholtz}\ \emph {et~al.}(2006)\citenamefont {Bergholtz}, \citenamefont {Kailasvuori}, \citenamefont {Wikberg}, \citenamefont {Hansson},\ and\ \citenamefont {Karlhede}}]{Bergholtz2006}%
  \BibitemOpen
  \bibfield  {author} {\bibinfo {author} {\bibfnamefont {E.~J.}\ \bibnamefont {Bergholtz}}, \bibinfo {author} {\bibfnamefont {J.}~\bibnamefont {Kailasvuori}}, \bibinfo {author} {\bibfnamefont {E.}~\bibnamefont {Wikberg}}, \bibinfo {author} {\bibfnamefont {T.~H.}\ \bibnamefont {Hansson}},\ and\ \bibinfo {author} {\bibfnamefont {A.}~\bibnamefont {Karlhede}},\ }\bibfield  {title} {\bibinfo {title} {Pfaffian quantum {Hall} state made simple: Multiple vacua and domain walls on a thin torus},\ }\href {https://doi.org/10.1103/PhysRevB.74.081308} {\bibfield  {journal} {\bibinfo  {journal} {Phys. Rev. B}\ }\textbf {\bibinfo {volume} {74}},\ \bibinfo {pages} {081308} (\bibinfo {year} {2006})}\BibitemShut {NoStop}%
\bibitem [{\citenamefont {Seidel}\ and\ \citenamefont {Lee}(2006)}]{Seidel2006}%
  \BibitemOpen
  \bibfield  {author} {\bibinfo {author} {\bibfnamefont {A.}~\bibnamefont {Seidel}}\ and\ \bibinfo {author} {\bibfnamefont {D.-H.}\ \bibnamefont {Lee}},\ }\bibfield  {title} {\bibinfo {title} {{Abelian} and non-{Abelian} {Hall} liquids and charge-density wave: Quantum number fractionalization in one and two dimensions},\ }\href {https://doi.org/10.1103/PhysRevLett.97.056804} {\bibfield  {journal} {\bibinfo  {journal} {Phys. Rev. Lett.}\ }\textbf {\bibinfo {volume} {97}},\ \bibinfo {pages} {056804} (\bibinfo {year} {2006})}\BibitemShut {NoStop}%
\bibitem [{\citenamefont {Su}\ and\ \citenamefont {MacDonald}(2008)}]{counterflow_macdonald}%
  \BibitemOpen
  \bibfield  {author} {\bibinfo {author} {\bibfnamefont {J.-J.}\ \bibnamefont {Su}}\ and\ \bibinfo {author} {\bibfnamefont {A.~H.}\ \bibnamefont {MacDonald}},\ }\bibfield  {title} {\bibinfo {title} {How to make a bilayer exciton condensate flow},\ }\href {https://doi.org/10.1038/nphys1055} {\bibfield  {journal} {\bibinfo  {journal} {Nature Physics}\ ,\ \bibinfo {pages} {799--802}} (\bibinfo {year} {2008})}\BibitemShut {NoStop}%
\bibitem [{\citenamefont {Zhang}\ \emph {et~al.}(2025)\citenamefont {Zhang}, \citenamefont {Nguyen}, \citenamefont {Batra}, \citenamefont {Liu}, \citenamefont {Watanabe}, \citenamefont {Taniguchi}, \citenamefont {Feldman},\ and\ \citenamefont {Li}}]{excitons_fqhe}%
  \BibitemOpen
  \bibfield  {author} {\bibinfo {author} {\bibfnamefont {N.~J.}\ \bibnamefont {Zhang}}, \bibinfo {author} {\bibfnamefont {R.~Q.}\ \bibnamefont {Nguyen}}, \bibinfo {author} {\bibfnamefont {N.}~\bibnamefont {Batra}}, \bibinfo {author} {\bibfnamefont {X.}~\bibnamefont {Liu}}, \bibinfo {author} {\bibfnamefont {K.}~\bibnamefont {Watanabe}}, \bibinfo {author} {\bibfnamefont {T.}~\bibnamefont {Taniguchi}}, \bibinfo {author} {\bibfnamefont {D.~E.}\ \bibnamefont {Feldman}},\ and\ \bibinfo {author} {\bibfnamefont {J.~I.~A.}\ \bibnamefont {Li}},\ }\bibfield  {title} {\bibinfo {title} {Excitons in the fractional quantum {Hall} effect},\ }\href {https://doi.org/10.1038/s41586-024-08274-3} {\bibfield  {journal} {\bibinfo  {journal} {Nature}\ }\textbf {\bibinfo {volume} {637}},\ \bibinfo {pages} {327--332} (\bibinfo {year} {2025})}\BibitemShut {NoStop}%
\bibitem [{\citenamefont {Unuchek}\ \emph {et~al.}(2018)\citenamefont {Unuchek}, \citenamefont {Ciarrocchi}, \citenamefont {Avsar}, \citenamefont {Watanabe}, \citenamefont {Taniguchi},\ and\ \citenamefont {Kis}}]{kis_exciton_transistor}%
  \BibitemOpen
  \bibfield  {author} {\bibinfo {author} {\bibfnamefont {D.}~\bibnamefont {Unuchek}}, \bibinfo {author} {\bibfnamefont {A.}~\bibnamefont {Ciarrocchi}}, \bibinfo {author} {\bibfnamefont {A.}~\bibnamefont {Avsar}}, \bibinfo {author} {\bibfnamefont {K.}~\bibnamefont {Watanabe}}, \bibinfo {author} {\bibfnamefont {T.}~\bibnamefont {Taniguchi}},\ and\ \bibinfo {author} {\bibfnamefont {A.}~\bibnamefont {Kis}},\ }\bibfield  {title} {\bibinfo {title} {Room-temperature electrical control of exciton flux in a van der {Waals} heterostructure},\ }\href {https://doi.org/10.1038/s41586-018-0357-y} {\bibfield  {journal} {\bibinfo  {journal} {Nature}\ }\textbf {\bibinfo {volume} {560}},\ \bibinfo {pages} {340--344} (\bibinfo {year} {2018})}\BibitemShut {NoStop}%
\bibitem [{\citenamefont {Rosati}\ \emph {et~al.}(2021)\citenamefont {Rosati}, \citenamefont {Schmidt}, \citenamefont {Brem}, \citenamefont {Perea-Causín}, \citenamefont {Niehues}, \citenamefont {Kern}, \citenamefont {Preuß}, \citenamefont {Schneider}, \citenamefont {de~Vasconcellos},\ and\ \citenamefont {Bratschitsch}}]{roberto_strain}%
  \BibitemOpen
  \bibfield  {author} {\bibinfo {author} {\bibfnamefont {R.}~\bibnamefont {Rosati}}, \bibinfo {author} {\bibfnamefont {R.}~\bibnamefont {Schmidt}}, \bibinfo {author} {\bibfnamefont {S.}~\bibnamefont {Brem}}, \bibinfo {author} {\bibfnamefont {R.}~\bibnamefont {Perea-Causín}}, \bibinfo {author} {\bibfnamefont {I.}~\bibnamefont {Niehues}}, \bibinfo {author} {\bibfnamefont {J.}~\bibnamefont {Kern}}, \bibinfo {author} {\bibfnamefont {J.~A.}\ \bibnamefont {Preuß}}, \bibinfo {author} {\bibfnamefont {R.}~\bibnamefont {Schneider}}, \bibinfo {author} {\bibfnamefont {S.~M.}\ \bibnamefont {de~Vasconcellos}},\ and\ \bibinfo {author} {\bibfnamefont {R.}~\bibnamefont {Bratschitsch}},\ }\bibfield  {title} {\bibinfo {title} {Dark exciton anti-funneling in atomically thin semiconductors},\ }\href {https://doi.org/10.1038/s41467-021-27425-y} {\bibfield  {journal} {\bibinfo  {journal} {Nature Communications}\ }\textbf {\bibinfo {volume} {12}},\ \bibinfo {pages} {7221} (\bibinfo {year} {2021})}\BibitemShut {NoStop}%
\bibitem [{\citenamefont {Tüğen}\ \emph {et~al.}(2025)\citenamefont {Tüğen}, \citenamefont {Seiler}, \citenamefont {Watanabe}, \citenamefont {Taniguchi}, \citenamefont {Kroner},\ and\ \citenamefont {İmamoğlu}}]{tugen2025opticalinjection}%
  \BibitemOpen
  \bibfield  {author} {\bibinfo {author} {\bibfnamefont {A.}~\bibnamefont {Tüğen}}, \bibinfo {author} {\bibfnamefont {A.~M.}\ \bibnamefont {Seiler}}, \bibinfo {author} {\bibfnamefont {K.}~\bibnamefont {Watanabe}}, \bibinfo {author} {\bibfnamefont {T.}~\bibnamefont {Taniguchi}}, \bibinfo {author} {\bibfnamefont {M.}~\bibnamefont {Kroner}},\ and\ \bibinfo {author} {\bibfnamefont {A.}~\bibnamefont {İmamoğlu}},\ }\href {https://arxiv.org/abs/2506.06098} {\bibinfo {title} {Optical injection and detection of long-lived interlayer excitons in van der {Waals} heterostructures}} (\bibinfo {year} {2025}),\ \Eprint {https://arxiv.org/abs/2506.06098} {arXiv:2506.06098 [cond-mat.mes-hall]} \BibitemShut {NoStop}%
\bibitem [{\citenamefont {Read}\ and\ \citenamefont {Rezayi}(1999)}]{read_rezayi}%
  \BibitemOpen
  \bibfield  {author} {\bibinfo {author} {\bibfnamefont {N.}~\bibnamefont {Read}}\ and\ \bibinfo {author} {\bibfnamefont {E.}~\bibnamefont {Rezayi}},\ }\bibfield  {title} {\bibinfo {title} {Beyond paired quantum {Hall} states: Parafermions and incompressible states in the first excited {Landau} level},\ }\href {https://doi.org/10.1103/PhysRevB.59.8084} {\bibfield  {journal} {\bibinfo  {journal} {Phys. Rev. B}\ }\textbf {\bibinfo {volume} {59}},\ \bibinfo {pages} {8084--8092} (\bibinfo {year} {1999})}\BibitemShut {NoStop}%
\bibitem [{\citenamefont {Fisher}\ \emph {et~al.}(1989)\citenamefont {Fisher}, \citenamefont {Weichman}, \citenamefont {Grinstein},\ and\ \citenamefont {Fisher}}]{fisher_superfluid_insulator}%
  \BibitemOpen
  \bibfield  {author} {\bibinfo {author} {\bibfnamefont {M.~P.~A.}\ \bibnamefont {Fisher}}, \bibinfo {author} {\bibfnamefont {P.~B.}\ \bibnamefont {Weichman}}, \bibinfo {author} {\bibfnamefont {G.}~\bibnamefont {Grinstein}},\ and\ \bibinfo {author} {\bibfnamefont {D.~S.}\ \bibnamefont {Fisher}},\ }\bibfield  {title} {\bibinfo {title} {Boson localization and the superfluid-insulator transition},\ }\href {https://doi.org/10.1103/PhysRevB.40.546} {\bibfield  {journal} {\bibinfo  {journal} {Phys. Rev. B}\ }\textbf {\bibinfo {volume} {40}},\ \bibinfo {pages} {546--570} (\bibinfo {year} {1989})}\BibitemShut {NoStop}%
\bibitem [{\citenamefont {Lu}\ \emph {et~al.}(2025)\citenamefont {Lu}, \citenamefont {Wu}, \citenamefont {Chen},\ and\ \citenamefont {Meng}}]{Lu_bosonic_superflied_transition}%
  \BibitemOpen
  \bibfield  {author} {\bibinfo {author} {\bibfnamefont {H.}~\bibnamefont {Lu}}, \bibinfo {author} {\bibfnamefont {H.-Q.}\ \bibnamefont {Wu}}, \bibinfo {author} {\bibfnamefont {B.-B.}\ \bibnamefont {Chen}},\ and\ \bibinfo {author} {\bibfnamefont {Z.~Y.}\ \bibnamefont {Meng}},\ }\bibfield  {title} {\bibinfo {title} {Continuous transition between bosonic fractional {Chern} insulator and superfluid},\ }\href {https://doi.org/10.1103/PhysRevLett.134.076601} {\bibfield  {journal} {\bibinfo  {journal} {Phys. Rev. Lett.}\ }\textbf {\bibinfo {volume} {134}},\ \bibinfo {pages} {076601} (\bibinfo {year} {2025})}\BibitemShut {NoStop}%
\bibitem [{\citenamefont {Huang}\ \emph {et~al.}(2024)\citenamefont {Huang}, \citenamefont {Lunts},\ and\ \citenamefont {Hafezi}}]{nonbosonic_excitons}%
  \BibitemOpen
  \bibfield  {author} {\bibinfo {author} {\bibfnamefont {T.-S.}\ \bibnamefont {Huang}}, \bibinfo {author} {\bibfnamefont {P.}~\bibnamefont {Lunts}},\ and\ \bibinfo {author} {\bibfnamefont {M.}~\bibnamefont {Hafezi}},\ }\bibfield  {title} {\bibinfo {title} {Nonbosonic moir\'e excitons},\ }\href {https://doi.org/10.1103/PhysRevLett.132.186202} {\bibfield  {journal} {\bibinfo  {journal} {Phys. Rev. Lett.}\ }\textbf {\bibinfo {volume} {132}},\ \bibinfo {pages} {186202} (\bibinfo {year} {2024})}\BibitemShut {NoStop}%
\bibitem [{\citenamefont {Schwartz}\ \emph {et~al.}(2021)\citenamefont {Schwartz}, \citenamefont {Shimazaki}, \citenamefont {Kuhlenkamp}, \citenamefont {Watanabe}, \citenamefont {Taniguchi}, \citenamefont {Kroner},\ and\ \citenamefont {Imamoğlu}}]{imamoglu_feshbach}%
  \BibitemOpen
  \bibfield  {author} {\bibinfo {author} {\bibfnamefont {I.}~\bibnamefont {Schwartz}}, \bibinfo {author} {\bibfnamefont {Y.}~\bibnamefont {Shimazaki}}, \bibinfo {author} {\bibfnamefont {C.}~\bibnamefont {Kuhlenkamp}}, \bibinfo {author} {\bibfnamefont {K.}~\bibnamefont {Watanabe}}, \bibinfo {author} {\bibfnamefont {T.}~\bibnamefont {Taniguchi}}, \bibinfo {author} {\bibfnamefont {M.}~\bibnamefont {Kroner}},\ and\ \bibinfo {author} {\bibfnamefont {A.}~\bibnamefont {Imamoğlu}},\ }\bibfield  {title} {\bibinfo {title} {Electrically tunable {Feshbach} resonances in twisted bilayer semiconductors},\ }\href {https://doi.org/10.1126/science.abj3831} {\bibfield  {journal} {\bibinfo  {journal} {Science}\ }\textbf {\bibinfo {volume} {374}},\ \bibinfo {pages} {336--340} (\bibinfo {year} {2021})}\BibitemShut {NoStop}%
\bibitem [{\citenamefont {Venanzi}\ \emph {et~al.}(2024)\citenamefont {Venanzi}, \citenamefont {Cuccu}, \citenamefont {Perea-Causin}, \citenamefont {Sun}, \citenamefont {Brem}, \citenamefont {Erkensten}, \citenamefont {Taniguchi}, \citenamefont {Watanabe}, \citenamefont {Malic}, \citenamefont {Helm}, \citenamefont {Winnerl},\ and\ \citenamefont {Chernikov}}]{chernikov_THz_trions}%
  \BibitemOpen
  \bibfield  {author} {\bibinfo {author} {\bibfnamefont {T.}~\bibnamefont {Venanzi}}, \bibinfo {author} {\bibfnamefont {M.}~\bibnamefont {Cuccu}}, \bibinfo {author} {\bibfnamefont {R.}~\bibnamefont {Perea-Causin}}, \bibinfo {author} {\bibfnamefont {X.}~\bibnamefont {Sun}}, \bibinfo {author} {\bibfnamefont {S.}~\bibnamefont {Brem}}, \bibinfo {author} {\bibfnamefont {D.}~\bibnamefont {Erkensten}}, \bibinfo {author} {\bibfnamefont {T.}~\bibnamefont {Taniguchi}}, \bibinfo {author} {\bibfnamefont {K.}~\bibnamefont {Watanabe}}, \bibinfo {author} {\bibfnamefont {E.}~\bibnamefont {Malic}}, \bibinfo {author} {\bibfnamefont {M.}~\bibnamefont {Helm}}, \bibinfo {author} {\bibfnamefont {S.}~\bibnamefont {Winnerl}},\ and\ \bibinfo {author} {\bibfnamefont {A.}~\bibnamefont {Chernikov}},\ }\bibfield  {title} {\bibinfo {title} {Ultrafast switching of trions in {2D} materials by terahertz photons},\ }\href {https://doi.org/10.1038/s41566-024-01512-0} {\bibfield  {journal} {\bibinfo  {journal} {Nature Photonics}\ ,\ \bibinfo
  {pages} {1344--1349}} (\bibinfo {year} {2024})}\BibitemShut {NoStop}%
\bibitem [{\citenamefont {Xie}\ \emph {et~al.}(2023)\citenamefont {Xie}, \citenamefont {Pan}, \citenamefont {Wu},\ and\ \citenamefont {Das~Sarma}}]{DasSarma_exciton_prl}%
  \BibitemOpen
  \bibfield  {author} {\bibinfo {author} {\bibfnamefont {M.}~\bibnamefont {Xie}}, \bibinfo {author} {\bibfnamefont {H.}~\bibnamefont {Pan}}, \bibinfo {author} {\bibfnamefont {F.}~\bibnamefont {Wu}},\ and\ \bibinfo {author} {\bibfnamefont {S.}~\bibnamefont {Das~Sarma}},\ }\bibfield  {title} {\bibinfo {title} {Nematic excitonic insulator in transition metal dichalcogenide moir\'e heterobilayers},\ }\href {https://doi.org/10.1103/PhysRevLett.131.046402} {\bibfield  {journal} {\bibinfo  {journal} {Phys. Rev. Lett.}\ }\textbf {\bibinfo {volume} {131}},\ \bibinfo {pages} {046402} (\bibinfo {year} {2023})}\BibitemShut {NoStop}%
\end{thebibliography}

%apsrev4-2.bst 2019-01-14 (MD) hand-edited version of apsrev4-1.bst
%Control: key (0)
%Control: author (8) initials jnrlst
%Control: editor formatted (1) identically to author
%Control: production of article title (0) allowed
%Control: page (1) range
%Control: year (1) truncated
%Control: production of eprint (0) enabled
%

\end{document}